\documentclass[journal,twoside]{IEEEtran}
\usepackage{amsmath,amsfonts}
\usepackage{mathrsfs}
\usepackage{tabularx}
\usepackage{algorithm}
\usepackage{algorithmic}
\usepackage{authblk}
\usepackage{hyperref}
\usepackage{xcolor}
\usepackage{doi}
\usepackage{array}
\usepackage[caption=false,font=normalsize,labelfont=sf,textfont=sf]{subfig}
\usepackage{textcomp}
\usepackage{stfloats}
\usepackage{url}
\usepackage{verbatim}
\usepackage{booktabs}
\usepackage{multirow}
\usepackage{makecell}
\usepackage{bm}
\usepackage{graphicx}
\usepackage{svg}
\usepackage{placeins}
\graphicspath{{Figures/PDF/}{Figures/PNG/}}
\usepackage[numbers,compress]{natbib}
\usepackage{orcidlink}
\def\BibTeX{{\rm B\kern-.05em{\sc i\kern-.025em b}\kern-.08em
		T\kern-.1667em\lower.7ex\hbox{E}\kern-.125emX}}
\usepackage{balance}

\begin{document}	
\bstctlcite{IEEEexample:BSTcontrol}

\title{Intrinsic Scatterer Representation for Forward Scattering Modeling of Complex Radar Targets}

\author{Ziyu Yue,~\IEEEmembership{Graduate Student Member,~IEEE,} and Feng Xu\raisebox{0.5ex}{\orcidlink{0000-0002-7015-1467}},~\IEEEmembership{Senior Member,~IEEE}
\thanks{This work has been submitted to the IEEE for possible publication.
	Copyright may be transferred without notice, after which this version may
	no longer be accessible.}
\thanks{
	The authors are with the Key Laboratory of Information Science of Electromagnetic Waves (Ministry of Education), Fudan University, Shanghai 200433, China (e-mail: fengxu@fudan.edu.cn).
}
}

\markboth{Journal of \LaTeX\ Class Files}%
{Yue \MakeLowercase{\textit{et al.}}: Intrinsic Scatterer Representation for Forward Scattering Modeling of Complex Radar Targets}

\maketitle

\begin{abstract}
	Forward modeling of scattering centers of radar targets is critical for advanced information retrieval of Synthetic Aperture Radar (SAR) images. 
	Existing forward modeling approaches rely on meshing the target and computing scattering via ray-tracing techniques, which not only incur high computational cost but also discard the semantic information of target geometry, thereby limiting the interpretability of SAR imagery.
	To address these issues, this paper proposes a novel forward scattering modeling framework that directly constructs stable, compact, and physically meaningful scatterers from target geometry.
	This formulation decouples target representation from specific observation configurations, enabling an intrinsic scatterer description across varying viewing angles.
	Specifically, an improved Random Sample Consensus (RANSAC) scheme is first developed to robustly extract planes, cylinders, and spheres from target point clouds, yielding single-bounce scatterers. 
	Potential multiple-bounce scatterers are then automatically detected by analyzing inter-primitive relations, and further refined through geometric clipping and parameter alignment to produce unique scattering representations compatible with canonical scattering center models. 
	Finally, radar responses and SAR images can be generated under arbitrary observation configurations based on the constructed scatterers.
	The proposed method is validated through extensive experiments, including comparisons with electromagnetic simulations and measured data. The results show that it can accurately characterize the scattering behaviors of complex targets, providing an efficient, interpretable, and reliable solution for SAR image modeling and understanding.
	
\end{abstract}



\section{Introduction}
\IEEEPARstart{S}{ynthetic} aperture radar (SAR) can acquire high-resolution microwave images of targets and environments from a distance regardless of weather and daylight, making it an essential tool for Earth remote sensing~\cite{xu2016preliminary}. High-resolution SAR images implicitly encode rich three-dimensional (3D) spatial cues, and interpreting these cues into human-understandable representations is the ultimate goal of advanced information retrieval (AIR)~\cite{xu2016preliminary} of SAR imagery. However, due to the inherent complexity of electromagnetic (EM) scattering phenomena and the unique imaging mechanisms of radar sensors, SAR images often exhibit sparse, point-like scattering patterns and high sensitivity to observation configurations, making their interpretation more challenging. Existing SAR image interpretation approaches can generally be categorized into model-free and model-based methods~\cite{xu2018emergence,feng2024microwave}. 
Model-free methods typically rely on expert-designed rules or large-size training data and often suffer from high annotation cost, long processing pipelines, or low robustness. 
Recent intelligent SAR image generation methods~\cite{xie2026azi,guan2026sarvehicle} can alleviate data scarcity by generating diverse samples with azimuth-dependent and scattering-related characteristics. 
Their primary focus, however, is image-level generation rather than explicitly establishing the correspondence between target 3D geometry and EM scattering mechanisms.
In contrast, model-based methods incorporate forward EM scattering models and leverage physical constraints to achieve stronger generalization capability and interpretability, making this paradigm a more promising research direction.

Apparently, a thorough understanding of the physical mechanisms of SAR imaging and the development of effective forward scattering models are indispensable for SAR AIR. 
An ideal forward scattering model should both support efficient forward simulations and be friendly to inverse problems~\cite{Xu2025SemanticScatteringCJORS}:
(1) low computational complexity and a high degree of automation, so as to accommodate flexible and large-scale data simulation;
(2) a compact and semantic representation with clear physical meaning, so as to reduce the difficulty of solving the associated inverse problems;
(3) a scattering computation paradigm that operates directly on such semantic entities, preserving physical interpretability and semantic consistency across both forward modeling and associated inverse analysis.

\vspace{0.2em}
\noindent\emph{A. Numerical and high-frequency methods lack semantic representation capability.}
\vspace{0.2em}

In existing studies on EM scattering modeling, numerical methods~\cite{Coifman1993FMM, Zhao2018SparseRoughSurface, Fu2016FEMCircuitCoupled} in computational electromagnetics compute scattered fields directly by discretizing Maxwell’s equations or their integral formulations. Although these methods provide high-accuracy solutions, they are generally impractical due to the prohibitive computational cost for large-scale and geometrically complex targets at radar-imaging scales. 

By contrast, high-frequency approximation methods~\cite{Keller1962GTD, Ufimtsev1962EdgeWaves, Knott1985ILDC, Roedder1999CADDSCAT} constitute the main tools for scattering modeling of radar targets. Built upon ray propagation with physical corrections, they exploit target geometry and reflection mechanisms to efficiently and accurately predict physical quantities such as the radar cross section (RCS). 
Despite avoiding direct field equation solving, these methods still rely on surface meshing of complex targets to determine ray interactions. 
The semantic information of targets is lost during mesh discretization: instead of meaningful structural entities such as “wings,” “nose,” or “engines,” the basic modeling units become physically uninformative facets, which fail to provide interpretable structural support for subsequent SAR image understanding.

\vspace{0.2em}
\noindent\emph{B. Scattering center modeling offers inherent semantic potential that remains largely untapped.}
\vspace{0.2em}

Scattering center modeling~\cite{Potter1997ASC, Gerry1999ParametricSAR, Jackson2010Canonical3DBistatic, Xing2021BistaticASC}, developed under high-frequency approximations, shows a semantic perspective by representing radar images as a coherent superposition of localized scattering centers associated with target structures. Each scattering center is described by a small number of physically meaningful parameters, such as location and size.

Most existing studies follow an inverse paradigm~\cite{Xing2022ScatteringReviewJR}, where scattering center parameters are estimated from observed SAR images. Another conceptually appealing direction is forward scattering center modeling~\cite{He2014ForwardASC, He2023ForwardTGRS, Zhu2025DielectricPECASC, Liu2025GeometrySC, zhang2025scattering}, which aims to derive equivalent scattering center representations directly from target geometry. 
Early work by He \emph{et al.}~\cite{He2014ForwardASC} proposed a forward modeling framework, which was subsequently extended to rough surfaces~\cite{He2023ForwardTGRS} and dielectric–PEC composite targets~\cite{Zhu2025DielectricPECASC}. 
From the perspective of semantic scattering modeling, such a direction is important because it seeks to convert geometric target representations, such as meshes, into scattering elements with semantic meaning. 

However, existing forward modeling studies still follow conventional procedures. 
Targets are first discretized into meshes, after which ray tracing is performed under given observation conditions to compute ray paths and scattered fields, extract dominant scattering sources, and further infer scattering center parameters. 
This discards the intrinsic geometric semantics of targets at the outset and derives a simplified scattering center representation only \emph{a posteriori} from conventional EM simulation results. 
This leads to three limitations: dependence on discretized meshes, limited semantic consistency during scattering computation, and complicated parameter inference based on traced-ray interactions. They also require handcrafted CAD modeling and component-level decomposition, making the workflow time-consuming and difficult to scale up.

Recent geometry-based studies~\cite{Liu2025GeometrySC,zhang2025scattering} have significantly advanced automatic forward modeling by extracting scatterers from meshed target geometry. In these methods, the observation configuration is introduced when determining illuminated and effective reflection regions, while multiple-bounce scatterers are identified through local reflection-path calculations between mesh facets. 
A complementary direction is therefore to decouple scatterer construction from observation-dependent response evaluation, allowing the same geometry-derived representation to be reused across viewing conditions.

\vspace{0.2em}
\noindent\emph{C. A new geometry-driven scattering center modeling framework enables fully semantic, tracing-free modeling.}
\vspace{0.2em}	

\begin{figure*}[t]
	\centering
	\includegraphics[width=1\linewidth]{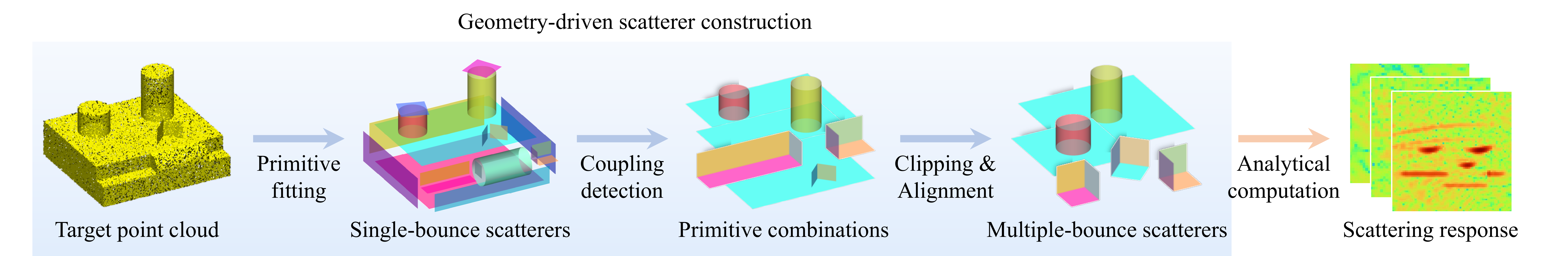}
	\caption{Overall procedure of the proposed forward modeling framework. The target point cloud is converted into an observation-independent scatterer representation (blue blocks), which is used with scattering center models~\cite{Jackson2010Canonical3DBistatic} to compute responses under arbitrary observation configurations (orange arrow).}
	\label{fig:pipeline}
\end{figure*}

In this paper, we propose a novel forward scattering center modeling framework, as illustrated in Fig.~\ref{fig:pipeline}. Its core idea is to derive scatterers directly from target geometry through primitive fitting (to identify single-bounce scatterers) and inter-primitive relationship analysis (to identify multiple-bounce scatterers), thereby mapping a target into a set of parameterized scatterers characterized by their location, size, and orientation. Given specific observation conditions, these geometric parameters can be directly fed into scattering center models to compute responses and simulate SAR images.

The derivation of scatterers is independent of radar observation conditions such as incidence angle and resolution. 
Unlike existing approaches, which must repeat scattering source extraction whenever observation conditions change, the scatterers obtained in our framework form a target-intrinsic representation and can be reused across different observation conditions.
Under this formulation, scattering computation no longer depends on facet-level meshes but operates directly on higher-level geometric structures, ensuring that the entire modeling process is conducted at the level of semantic entities.

To realize this framework, inspired by primitive fitting techniques in CAD reverse engineering~\cite{Schnabel2007EfficientRANSAC,Li2011Globfit,Li2023SurfaceEdge,Liu2024SplitAndFit}, we propose an improved Random Sample Consensus (RANSAC)~\cite{Schnabel2007EfficientRANSAC} algorithm tailored for scatterer extraction. 
The proposed method automatically identifies planar, cylindrical, and spherical primitives from target point clouds, which serve as candidates for single-scattering searching. 
Subsequently, a set of directional and adjacency criteria is designed to determine whether different primitives jointly form multiple-scattering mechanisms, such as dihedrals, trihedrals, and top-hats. 
For each detected multiple-bounce scatterer, an effective-region clipping strategy is introduced to extract the physically valid scattering regions and to determine parameters compatible with canonical scattering center models proposed by Jackson \emph{et al.}~\cite{Jackson2010Canonical3DBistatic}. 
Finally, these parameters are fed into the models to compute target scattering responses under arbitrary observation conditions.

Experiments are conducted on multiple test cases. Comparative results against FEKO RL-GO EM simulations and measured data demonstrate the efficacy of the proposed method, with evaluations covering geometric parameter accuracy, SAR image similarity, peak matching, scattering-structure consistency, full-polarimetric response, and computational efficiency.

The main contributions are summarized as follows:

\begin{enumerate}
	\item  A novel forward scattering center modeling framework: A 3D representation of targets is constructed from their intrinsic geometry rather than from EM propagation paths, shifting the paradigm from \emph{view-dependent computation} to \emph{intrinsic target characterization}. It supports scattering modeling that operates at the semantic level and provides a representational foundation for SAR image understanding.
	\item An automatic method for scatterer construction: A unified procedure integrating geometric primitive detection, inter-primitive coupling analysis, and effective-region clipping is developed, which enables accurate construction of potential single- and multiple-bounce scatterers contained in targets, without manual intervention or ray tracing.
	\item Extensive demonstration and evaluation: Scattering modeling and SAR imaging simulations are conducted on multiple targets and evaluated from multiple aspects, showing high agreement with EM simulations and substantially improving computational efficiency.
	
\end{enumerate}

The remainder of this paper is organized as follows. Section~\ref{sec:single} describes the automatic fitting of single-scattering. Section~\ref{sec:multi} presents the detection of multiple-scattering and the determination of scattering parameters. Section~\ref{sec:exp} validates the proposed method through extensive experiments. Finally, Section~\ref{sec:conclusion} concludes the paper and discusses future work.


\section{RANSAC-Based Primitive Scatterer Fitting}
\label{sec:single}

In the high-frequency regime, the scattering response of a radar target can be regarded as the coherent superposition of a set of dominant scattering centers localized at specific regions. 
Based on high-frequency asymptotic scattering theory, Jackson \emph{et al.}~\cite{Jackson2010Canonical3DBistatic} constructed 3D bistatic scattering models for six canonical structures, including the plane, dihedral, trihedral, cylinder, top-hat, and sphere.

Let $f$ denote the radar frequency, $\Lambda$ the 3D bistatic observation configuration, and $\Theta$ the geometric parameters of a scatterer. Its scattering response is expressed as
\begin{equation}
	S(f,\Lambda;\Theta)
	=
	P(f,\Lambda;\Theta)\,
	M(f,\Lambda;\Theta)\,
	\exp\!\left(j\Phi(f,\Lambda;\Theta)\right),
	\label{eq:canonical_scatter}
\end{equation}
where $\Lambda=(\theta_i,\theta_s,\phi_i,\phi_s)$ denotes the incident and scattered elevation and azimuth angles. The canonical models are defined in a standard local frame, with $\Theta$ specifying intrinsic position and size, while orientation is handled by transforming the global observation directions into this local frame.

In~\eqref{eq:canonical_scatter}, $P(\cdot)$ describes the polarization dependence, $M(\cdot)$ denotes the structure-specific shape function, and $\Phi(\cdot)$ models the phase. Complete expressions can be found in~\cite{Jackson2010Canonical3DBistatic}.

\subsection{Overview of Primitive Fitting}

\begin{figure}[t]
	\centering
	\includegraphics[width=0.8\linewidth]{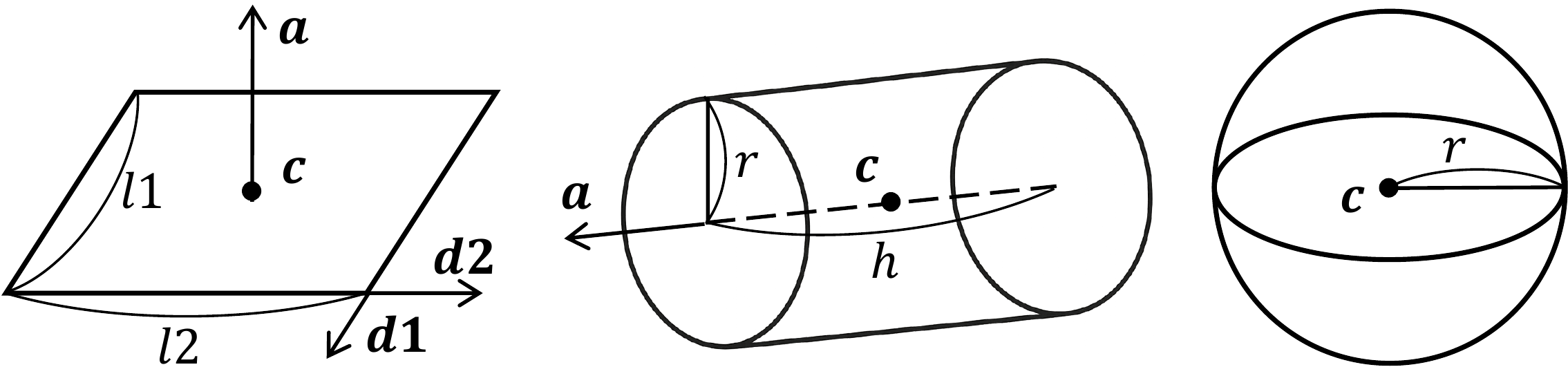}
	\caption{Geometric primitives and their parameters used in this work.}
	\label{fig:primitive_params}
\end{figure}

It can be observed that the six scatterers can be fully described using three irreducible geometric primitives: the plane, cylinder, and sphere. The remaining three multiple-bounce structures are formed by combinations of these primitives. Accordingly, we define a primitive dictionary
\begin{equation}
	\mathcal{D} = \left\{
	\begin{aligned}
		&s_{\text{pl}} = \mathrm{Plane}(\mathbf{c},\mathbf{a},\mathbf{d}_1,\mathbf{d}_2,l_1,l_2),\\
		&s_{\text{cy}} = \mathrm{Cylinder}(\mathbf{c},\mathbf{a},r,h),\\
		&s_{\text{sp}} = \mathrm{Sphere}(\mathbf{c},r)
	\end{aligned}
	\right\},
	\label{eq:primitive_dictionary}
\end{equation}
where Fig.~\ref{fig:primitive_params} illustrates the parameters. Specifically, $\mathbf{c}$ denotes the primitive center, $\mathbf{a}$ denotes the plane normal or cylinder axis, $\mathbf{d}_1$ and $\mathbf{d}_2$ are the in-plane edge directions, and $l_1,l_2,r,h$ define the spatial extents of primitives.

Given an oriented point cloud
$\mathcal{P}=\{(\mathbf{p}_i,\mathbf{n}_i)\}_{i=0}^{N-1}$ with consistently outward-facing normals, we employ a RANSAC-based shape detection framework~\cite{Schnabel2007EfficientRANSAC} to recover a set of bounded, parameterized primitives
$\mathscr{S}=\{s_i\}_{i=0}^{M-1}$.
As summarized in Algorithm~\ref{alg:alg1}, geometric hypotheses are repeatedly generated from sampled point subsets, evaluated based on inlier support, and refined to extract primitives with clear physical meaning from cluttered point clouds.

\begin{figure}[t]
	\centering
	\includegraphics[width=0.8\linewidth]{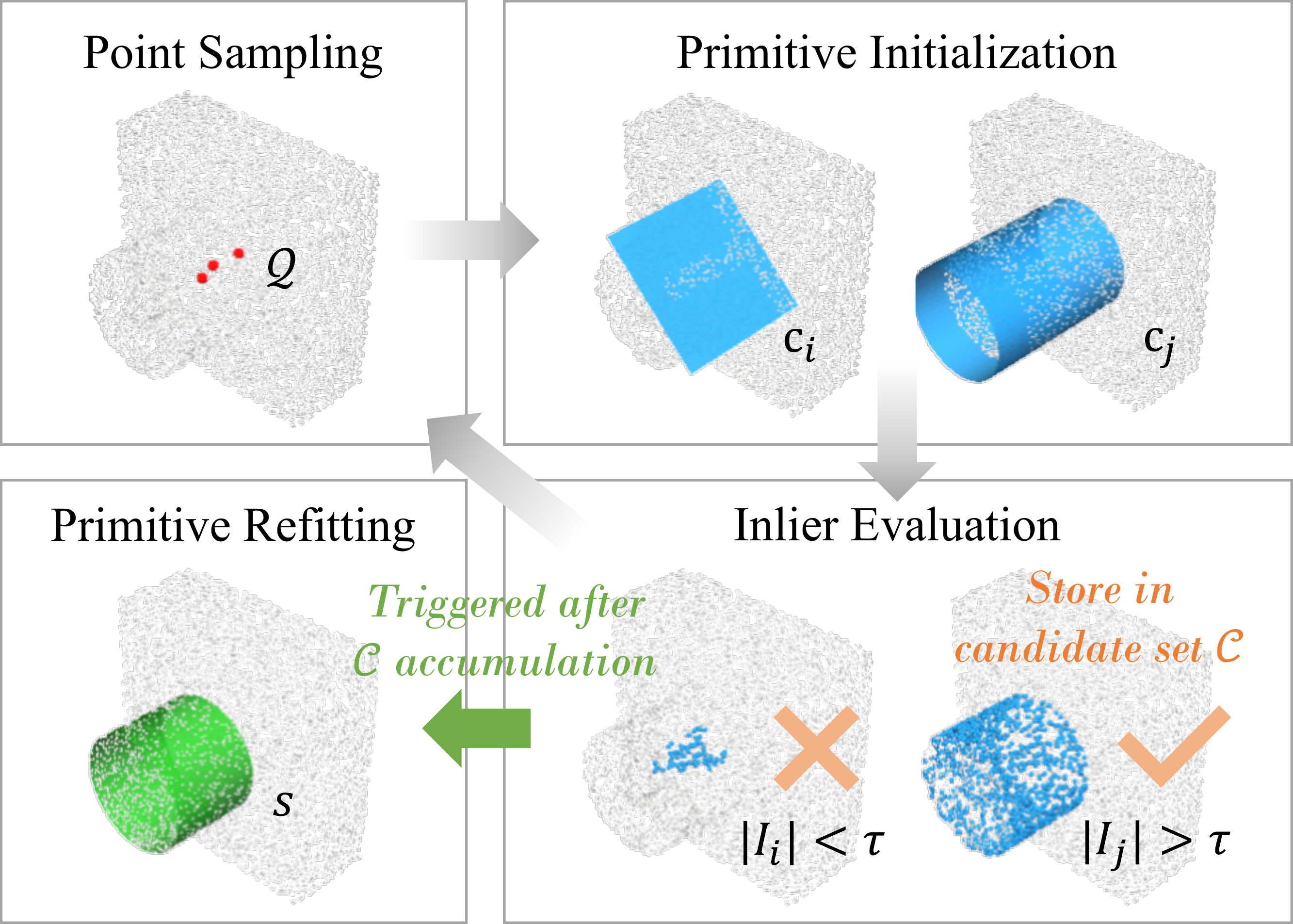}
	\caption{One RANSAC-based primitive-detection cycle. Minimal samples $\mathcal{Q}$ generate hypotheses $c_d$, and the best-supported candidate is refitted as the bounded primitive $s$. Only planes and cylinders are shown.}
	\label{fig:ransac_pipeline}
\end{figure}

Building on the original algorithm, we tailor the refitting stage to the requirements of canonical scatterer construction. For planar primitives, boundary estimation is reformulated as a minimum-area bounding rectangle problem in a projected 2D domain, avoiding the instability of principal component analysis (PCA) in estimating rectangle edge directions. For cylindrical primitives, instead of nonlinear joint optimization—which is prone to divergence—we adopt deterministic closed-form geometric solutions based on point-pair constraints to obtain stable radius and height estimates. The following subsections sequentially describe the hypothesis initialization, inlier evaluation, and primitive refitting.

\begin{algorithm}[t]
	\small
	\caption{{\small RANSAC-Based Primitive Scatterer Fitting}}
	\label{alg:alg1}
	\begin{algorithmic}[1]			
		\REQUIRE
		
		\begin{tabular}[t]{@{}l@{\;}c@{\hspace{3em}}l@{\;}c@{}}
			Oriented point cloud:
			& \multicolumn{3}{l}{
				$\mathcal{P}=\{(\mathbf{p}_i,\mathbf{n}_i)\}_{i=0}^{N-1}$
			}\\
			Primitive dictionary:
			& \multicolumn{3}{l}{
				$\mathcal{D}=\{\text{Plane},\text{Cylinder},\text{Sphere}\}$
			}\\
			Min. support:       & $\tau$
			& Distance threshold:   & $\varepsilon$\\
			Normal threshold:   & $\alpha$
			& Bitmap resolution:    & $\beta$\\
			Confidence threshold: & $\eta$
			& Max. iterations:      & $T$
		\end{tabular}
		
		\ENSURE Primitive set $\mathscr{S}$.
		
		\STATE Initialize $\mathscr{S}\leftarrow\emptyset$.
		
		\WHILE{$|\mathcal{P}| > \tau$}
		
		\STATE Initialize candidate set $\mathcal{C}\leftarrow\emptyset$, iteration counter $t\leftarrow0$.
		
		\FOR{$t=1,\ldots,T$}
		
		\STATE Sample a minimal point set $\mathcal{Q}\subset\mathcal{P}$.
		
		\FOR{each $d \in \mathcal{D}$}
		\STATE Initialize a hypothesis $c_d$ from $\mathcal{Q}$. (Sec.~\ref{sec:step1-1})
		\STATE Select inliers satisfying distances $d_i<\varepsilon$ and normal consistencies $\gamma_i>\alpha$. (Sec.~\ref{sec:step1-2})
		\STATE Rasterize with $\beta$ and retain the largest connected component $I_d$. (Sec.~\ref{sec:step1-2})
		\IF{$|I_d|>\tau$ and $(c_d,I_d)\notin \mathcal{C}$}
		\STATE $\mathcal{C}\leftarrow \mathcal{C}\cup\{(c_d,I_d)\}$.
		\ENDIF
		\ENDFOR

		\STATE Select the best candidate $(c^\ast,I^\ast)$. (Sec.~\ref{sec:step1-3})
		\IF{its confidence exceeds $\eta$}
		\STATE \textbf{break}
		\ENDIF

		\ENDFOR

		\STATE Refit $c^\ast$ using $I^\ast$ to obtain the final primitive $s$. (Sec.~\ref{sec:step1-3})
		\STATE $\mathscr{S}\leftarrow\mathscr{S}\cup\{s\}$;
		$\mathcal{P}\leftarrow\mathcal{P}\setminus I^\ast$
		
		\ENDWHILE
		
		\RETURN $\mathscr{S}$
		
	\end{algorithmic}
\end{algorithm}

\subsection{Primitive Initialization}
\label{sec:step1-1}

As illustrated in Algorithm~\ref{alg:alg1}, the algorithm first initializes several key parameters. 
The minimum support size $\tau$ specifies the minimum number of inliers required for a primitive to be accepted. At each iteration, the algorithm checks whether the number of remaining unassigned points is larger than $\tau$. If so, three points $\mathcal{Q}=\{\mathbf{p}_0, \mathbf{p}_1, \mathbf{p}_2\}$ together with their normals $\{\mathbf{n}_0,\mathbf{n}_1,\mathbf{n}_2\}$ are randomly sampled from the remaining set (red points in Fig.~\ref{fig:ransac_pipeline}). Using the three points, unbounded primitives $c_d$ are estimated for each primitive type based on shape-specific priors (blue surfaces in Fig.~\ref{fig:ransac_pipeline}). For planar and cylindrical primitives, the position parameter $\mathbf{c}$ is temporarily chosen as an arbitrary point on the plane or the axis; its accurate value, together with other bounded-support size parameters, will be determined during the refitting stage.

\paragraph{Plane}
Given three sampled points $\{\mathbf{p}_0,\mathbf{p}_1,\mathbf{p}_2\}$, the plane normal and center are obtained as
\begin{equation}
	\mathbf{a} =
	\frac{(\mathbf{p}_2-\mathbf{p}_1)\times(\mathbf{p}_1-\mathbf{p}_0)}
	{\|(\mathbf{p}_2-\mathbf{p}_1)\times(\mathbf{p}_1-\mathbf{p}_0)\|},
	\qquad
	\mathbf{c}=\mathbf{p}_0,
	\label{eq:step1_plane}
\end{equation}
yielding the infinite plane defined by $\mathbf{a}^\top(\mathbf{x}-\mathbf{c})=0$.

\paragraph{Cylinder}
A cylindrical hypothesis is initialized from two oriented points
$\{(\mathbf{p}_0,\mathbf{n}_0),(\mathbf{p}_1,\mathbf{n}_1)\}$. Its axis
direction is estimated as
\begin{equation}
	\mathbf{a}=
	\frac{\mathbf{n}_1\times\mathbf{n}_0}
	{\|\mathbf{n}_1\times\mathbf{n}_0\|}.
	\label{eq:cylinder_axis}
\end{equation}
To estimate the radius $r$ and a point $\mathbf{c}$ on the axis, all points are projected onto a plane perpendicular to $\mathbf{a}$. Enforcing that the two projected normals intersect at the cross-sectional center gives a closed-form estimate of the radius $r$ and an
axis point
\begin{equation}
	\mathbf{c}=\mathbf{p}_0-r\mathbf{n}_0.
	\label{eq:cylinder_center}
\end{equation}
The finite axial extent is determined later from the complete inlier set.

\paragraph{Sphere}
A spherical hypothesis is similarly initialized from two oriented points
$\{(\mathbf{p}_0,\mathbf{n}_0),(\mathbf{p}_1,\mathbf{n}_1)\}$. 
We parameterize two lines
\begin{equation}
	\mathbf{L}_0(\mu_0)=\mathbf{p}_0+\mu_0\mathbf{n}_0,\quad
	\mathbf{L}_1(\mu_1)=\mathbf{p}_1+\mu_1\mathbf{n}_1,
\end{equation}
and compute the sphere center as the midpoint of their shortest connecting segment:
\begin{equation}
	\begin{aligned}
		(\mu_0^{*},\mu_1^{*})
		&=
		\arg\min_{\mu_0,\mu_1}
		\|\mathbf{L}_0(\mu_0)-\mathbf{L}_1(\mu_1)\|^2,\\
		\mathbf{c}
		&=
		\frac{\mathbf{L}_0(\mu_0^{*})+
			\mathbf{L}_1(\mu_1^{*})}{2},\\
		r
		&=
		\frac{\|\mathbf{p}_0-\mathbf{c}\|+
			\|\mathbf{p}_1-\mathbf{c}\|}{2}.
	\end{aligned}
	\label{eq:sphere_initialization}
\end{equation}

\subsection{Inlier Evaluation}
\label{sec:step1-2}

For each hypothesis $c_d$, its support is evaluated on the current point cloud, i.e., the number of inliers $I_d$ (blue points in Fig.~\ref{fig:ransac_pipeline}). If $|I_d|$ exceeds $\tau$, the primitive $(c_d, I_d)$ is added to the candidate set $\mathcal{C}$. Three parameters are introduced for inlier selection: a distance threshold $\varepsilon$, a normal-consistency threshold $\alpha$, and a connectivity bitmap resolution $\beta$.

Given a point $(\mathbf{p}_i,\mathbf{n}_i)$, its distance $d_i$ to a primitive surface and the cosine of the angle $\gamma_i$ between its normal and the primitive normal are computed. The point is initially regarded as an inlier if it satisfies $d_i < \varepsilon$ and $\gamma_i > \alpha$.

\paragraph{Plane}
For a planar hypothesis $c_d=\mathrm{Plane}(\mathbf{c},\mathbf{a})$, the distance and normal consistency are given by
\begin{equation}
	d_i =
	\left|\mathbf{a}^{\top}(\mathbf{p}_i-\mathbf{c})\right|,
	\qquad
	\gamma_i =
	\left|\mathbf{n}_i^{\top}\mathbf{a}\right|.
	\label{eq:plane_inlier_metrics}
\end{equation}
The absolute value accounts for the sign ambiguity of the plane normal.

\paragraph{Cylinder}
For $c_d=\mathrm{Cylinder}(\mathbf{c},\mathbf{a},r)$, let
\begin{equation}
	\mathbf{v}_i = \mathbf{p}_i-\mathbf{c}-\big((\mathbf{p}_i-\mathbf{c})^{\top}\mathbf{a}\big)\mathbf{a}
\end{equation}
denote the radial component of $\mathbf{p}_i-\mathbf{c}$. 
The distance and normal consistency are computed as
\begin{equation}
	d_i =
	\left|\|\mathbf{v}_i\|-r\right|,
	\qquad
	\gamma_i =
	\frac{\mathbf{n}_i^{\top}\mathbf{v}_i}{\|\mathbf{v}_i\|}.
	\label{eq:cylinder_inlier_metrics}
\end{equation}

\paragraph{Sphere}
For a spherical hypothesis $c_d=\mathrm{Sphere}(\mathbf{c},r)$, the corresponding quantities are
\begin{equation}
	d_i =
	\left|\|\mathbf{p}_i-\mathbf{c}\|-r\right|,
	\qquad
	\gamma_i =
	\frac{\mathbf{n}_i^{\top}(\mathbf{p}_i-\mathbf{c})}
	{\|\mathbf{p}_i-\mathbf{c}\|}.
	\label{eq:sphere_inlier_metrics}
\end{equation}

Distance and normal consistency alone may merge spatially disconnected regions that satisfy the same primitive equation. Following \cite{Schnabel2007EfficientRANSAC}, the preliminary inliers are therefore mapped to a primitive-specific 2D parameter domain: planar coordinates for planes, angular--axial coordinates for cylinders, and spherical angles for spheres. The parameter domain is rasterized with resolution $\beta$, with periodic coordinates treated cyclically, and only the largest 8-connected component is retained as the final inlier set $I_d$.

A hypothesis is added to the candidate set $\mathcal{C}$ if $|I_d|>\tau$ and it is not geometrically equivalent to an existing
candidate.

\subsection{Primitive Refitting}
\label{sec:step1-3}

As candidate hypotheses accumulate, the primitive with the largest connected inlier support is periodically selected as the current optimum $(c^\ast,I^\ast)$. Its detection confidence is evaluated using the standard RANSAC criterion. Sampling terminates when the confidence exceeds threshold $\eta$ or the iteration count reaches $T$. The selected candidate is then refitted using all inliers in $I^\ast$, which both improves its parameter accuracy and determines its bounded support. The resulting fully parameterized primitive $s$ (green surface in Fig.~\ref{fig:ransac_pipeline}) is added to $\mathscr{S}$.

\paragraph{Plane}
For a planar primitive, PCA is first applied to the inlier coordinates to estimate the surface normal $\mathbf{a}$, whose sign is aligned with the mean inlier normal to ensure a consistent outward-facing orientation. 
The inliers are then projected onto a local 2D basis $(\mathbf{e}_1,\mathbf{e}_2)$ perpendicular to $\mathbf{a}$.
Unlike directly using the two principal directions as the rectangle axes, which can be unstable for near-isotropic in-plane point distributions, we determine the bounded support through the minimum-area bounding rectangle of the projected convex hull.

Let $\mathcal{H}=\operatorname{conv}\{\mathbf{q}_j\}_{j=1}^{N_c}$ denote the 2D convex hull, and let $\mathcal{E}(\mathcal{H})$ be the set of its unit edge directions. Define the extent of $\mathcal{H}$ along a direction $\mathbf{u}$
as
\begin{equation}
	\Delta_{\mathbf{u}}(\mathcal{H})
	=
	\max_j \mathbf{q}_j^\top\mathbf{u}
	-
	\min_j \mathbf{q}_j^\top\mathbf{u}.
\end{equation}
The rectangle axes are selected by
\begin{equation}
	\mathbf{d}_1^{2\mathrm{D}}
	=
	\arg\min_{\mathbf{u}\in\mathcal{E}(\mathcal{H})}
	\Delta_{\mathbf{u}}(\mathcal{H})
	\Delta_{\mathbf{u}^{\perp}}(\mathcal{H}),
	\qquad
	\mathbf{d}_2^{2\mathrm{D}}
	=
	\left(\mathbf{d}_1^{2\mathrm{D}}\right)^{\perp}.
	\label{eq:min_area_rect}
\end{equation}
The extrema along these two directions determine the rectangle center and side lengths $(l_1,l_2)$, which are subsequently lifted to the original 3D coordinate system through $(\mathbf{e}_1,\mathbf{e}_2)$. This yields the refitted plane $s=\mathrm{Plane}(\mathbf{c},\mathbf{a},\mathbf{d}_1,\mathbf{d}_2,l_1,l_2)$.

\paragraph{Cylinder}
A straightforward refitting approach would be to jointly optimize the cylinder axis, center, and radius using nonlinear least squares (e.g., Levenberg--Marquardt)~\cite{Schnabel2007EfficientRANSAC}. However, in practice such optimization is prone to divergence and often forces nearly planar regions to be incorrectly fitted as cylinders. We therefore continue to adopt a deterministic geometric refitting strategy.

PCA of the inlier normals is first used to estimate the axis $\mathbf{a}$. Multiple valid oriented point pairs are then used to generate closed-form estimates of the cross-sectional center and radius according to the point--normal geometry described in Sec.~\ref{sec:step1-1}. The final axis point $\mathbf{c}$ and radius $r$ are obtained by averaging the valid estimates. The finite axial extent is determined by projecting all inliers onto the estimated axis:
\begin{equation}
	h
	=
	\max_{(\mathbf{p}_i,\mathbf{n}_i)\in I^\ast}
	\mathbf{p}_i^\top\mathbf{a}
	-
	\min_{(\mathbf{p}_i,\mathbf{n}_i)\in I^\ast}
	\mathbf{p}_i^\top\mathbf{a}.
	\label{eq:cyl_height}
\end{equation}
The refitted primitive is therefore
$s=\mathrm{Cylinder}(\mathbf{c},\mathbf{a},r,h)$.

\paragraph{Sphere}
The center and radius are obtained as
\begin{equation}
	\begin{aligned}
		\mathbf{c}
		=
		\arg\min_{\mathbf{x}}
		&\sum_{(\mathbf{p}_i,\mathbf{n}_i)\in I^\ast}
		\left\|
		(\mathbf{I}_3-\mathbf{n}_i\mathbf{n}_i^\top)
		(\mathbf{x}-\mathbf{p}_i)
		\right\|^2,\\
		r
		=
		\frac{1}{|I^\ast|}
		&\sum_{(\mathbf{p}_i,\mathbf{n}_i)\in I^\ast}
		\|\mathbf{p}_i-\mathbf{c}\|,
	\end{aligned}
	\label{eq:sphere_refitting}
\end{equation}
yielding $s=\mathrm{Sphere}(\mathbf{c},r)$.

After an accepted primitive is added to $\mathscr{S}$, its inliers are removed
from $\mathcal{P}$, and the procedure is repeated on the remaining
points until they are insufficient to support new primitive hypotheses.


\section{Parameter Determination for Scatterer Models}
\label{sec:multi}

\begin{figure}[b]
	\centering
	\includegraphics[width=0.8\linewidth]{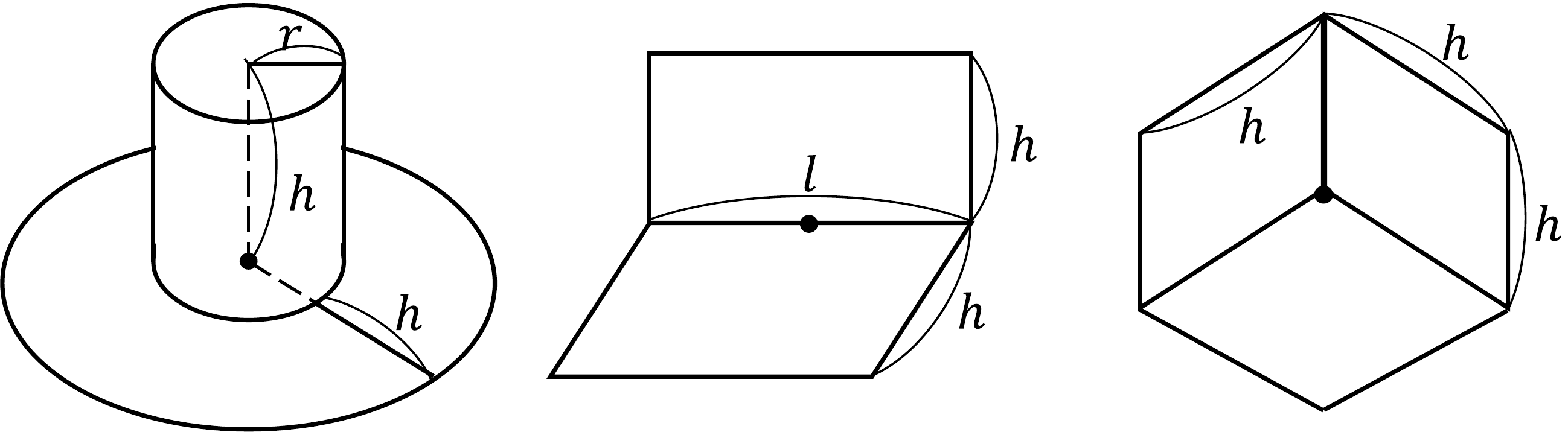}
	\caption{Multiple-bounce scatterers in the canonical scattering center models~\cite{Jackson2010Canonical3DBistatic}, including top-hat, dihedral, and trihedral structures.}
	\label{fig:multi_scatterers}
\end{figure}

Each primitive in $\mathscr{S}$ is treated as a candidate single-bounce scatterer. Potential dihedrals, trihedrals, and top-hats are then identified from geometric relations among multiple primitives (Fig.~\ref{fig:multi_scatterers}). Their effective regions are clipped where necessary, and the resulting parameters are aligned with the canonical scattering center models, as summarized in Algorithm~\ref{alg:alg2}.
In addition, Sec.~\ref{sec:global_scattering} briefly describes how the
constructed scatterers are used to synthesize the overall target response.

\subsection{Primitive Coupling Detection}

We first identify coupled primitive combinations that may give rise to multiple-scattering.

\paragraph{Dihedral (plane--plane)}
For dihedral detection, all pairs of planes $s_i$ and $s_j$ in $\mathscr{S}$ are examined. If both the directional constraint in~\eqref{eq:dihedral_conditions_a} and the adjacency constraint in~\eqref{eq:dihedral_conditions_b} are satisfied, the pair is then recorded as a potential dihedral $d_{ij}$ and added to the scatterer set $\mathscr{S}$.

\begin{figure}[t]
	\centering
	\includegraphics[width=0.8\linewidth]{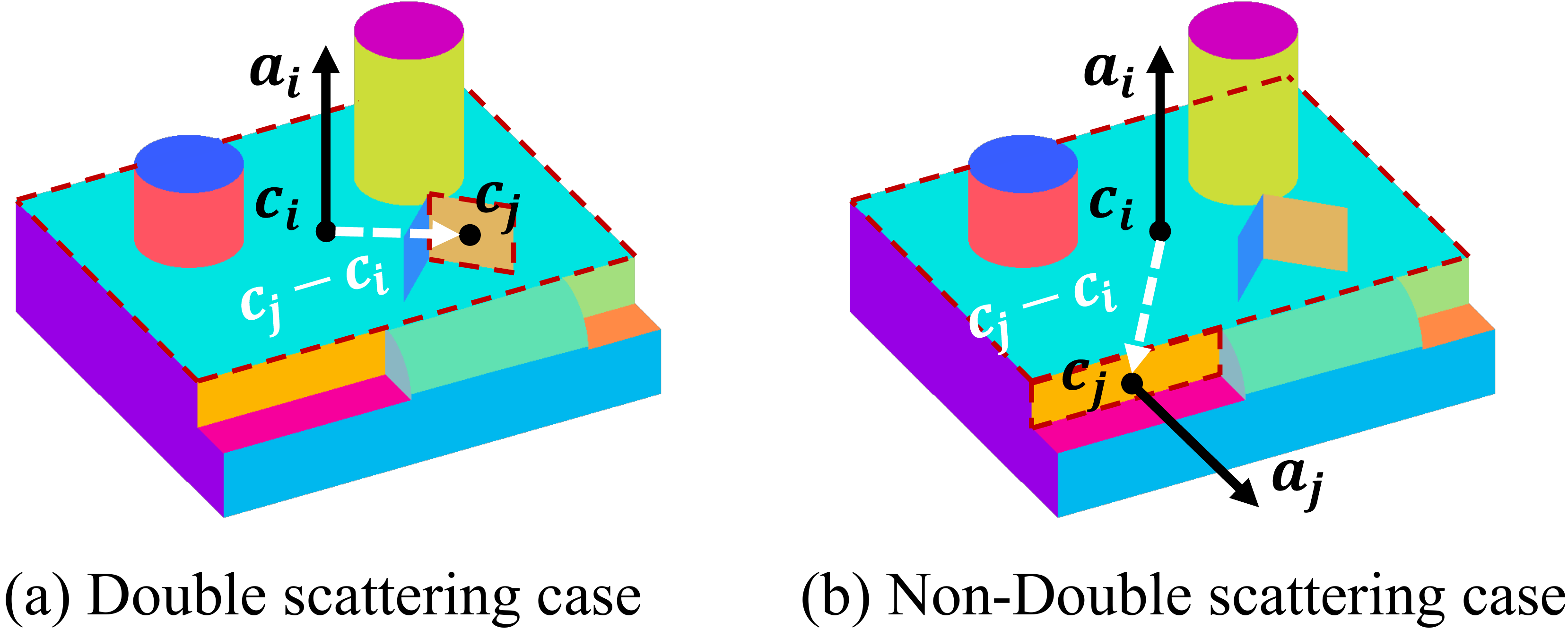}
	\caption{Directional criterion for dihedral detection. Red dashed boxes mark the evaluated plane pairs, with centers $\mathbf{c}$ and normals $\mathbf{a}$: (a) satisfied; (b) violated.}
	\label{fig:dihedral_rule}
\end{figure}

Fig.~\ref{fig:dihedral_rule} illustrates the directional criterion on the
MSTAR Sandia laboratory implementation of cylinders (SLICY) model
\cite{Diemunsch1998MSTARATR}. The adjacent planes in
Fig.~\ref{fig:dihedral_rule}(a) satisfy the criterion and are retained,
whereas those in Fig.~\ref{fig:dihedral_rule}(b) are excluded because their
relative orientation violates it.
\begin{subequations}\label{eq:dihedral_conditions}
	\begin{align}
		(\mathbf{c}_j-\mathbf{c}_i)\cdot\mathbf{a}_i \ge 0 
		\;\;\text{or}\;\;
		(\mathbf{c}_i-\mathbf{c}_j)\cdot\mathbf{a}_j \ge 0, \label{eq:dihedral_conditions_a}\\
		\min_{\mathbf{p}_i \in s_i,\;\mathbf{p}_j \in s_j}
		\|\mathbf{p}_i-\mathbf{p}_j\| < \delta. \label{eq:dihedral_conditions_b}
	\end{align}
\end{subequations}

\paragraph{Top-hat (plane--cylinder)}
All plane--cylinder pairs $(s_i, s_j)$ in $\mathscr{S}$ are examined. If both the directional constraint
\begin{equation}
	(\mathbf{c}_j-\mathbf{c}_i)\cdot\mathbf{a}_i \ge 0
	\label{eq:tophat_dir}
\end{equation}
and the adjacency constraint in~\eqref{eq:dihedral_conditions_b} are satisfied, the pair is regarded as a potential top-hat scatterer, denoted by $d_{ij}$, and is added to the scatterer set $\mathscr{S}$.

\paragraph{Trihedral (plane--plane--plane)}
If three planes satisfy that every two of them form a dihedral, i.e., $d_{ij}, d_{ik}, d_{jk} \in \mathscr{S}$, the triplet is regarded as a trihedral scatterer and recorded as $t_{ijk}$. The triplet  is then added to  $\mathscr{S}$.

Fig.~\ref{fig:slicy_multiples} shows the multiple-bounce scatterers detected on the SLICY target using the proposed geometric coupling rules.

\begin{algorithm}[t]
	\small
	\caption{{\small Multiple-Bounce Scatterer Construction}}
	\label{alg:alg2}
	\begin{algorithmic}[1]
		
		\REQUIRE 
		\begin{tabular}[t]{@{}ll@{}}
			Primitive set: & $\mathscr{S}=\{s_i\}$ (from Alg.~\ref{alg:alg1})\\
			Adjacency threshold: & $\delta$\\
		\end{tabular}
		
		\ENSURE Updated scatterer set $\mathscr{S}$.
		
		\vspace{0.1em}
		\STATE \textbf{(1) Primitive coupling detection}
		
		\FORALL{plane pairs $(s_i,s_j)\subset\mathscr{S}$}
		\IF{Eqs.~\eqref{eq:dihedral_conditions_a}--\eqref{eq:dihedral_conditions_b} satisfied}
		\STATE Construct dihedral scatterer $d_{ij}$. $\mathscr{S}\leftarrow\mathscr{S}\cup\{d_{ij}\}$.
		\ENDIF
		\ENDFOR
		
		\FORALL{plane--cylinder pairs $(s_i,s_j)\subset\mathscr{S}$}
		\IF{Eqs.~\eqref{eq:tophat_dir},~\eqref{eq:dihedral_conditions_b} satisfied}
		\STATE Construct top-hat scatterer $d_{ij}$. $\mathscr{S}\leftarrow\mathscr{S}\cup\{d_{ij}\}$.
		\ENDIF
		\ENDFOR
		
		\FORALL{plane triplets $(s_i,s_j,s_k)\subset\mathscr{S}$}
		\IF{$d_{ij},d_{ik},d_{jk}\in\mathscr{S}$}
		\STATE Construct trihedral scatterer $t_{ijk}$. $\mathscr{S}\leftarrow\mathscr{S}\cup\{t_{ijk}\}$.
		\ENDIF
		\ENDFOR
		
		\vspace{0.3em}
		\STATE \textbf{(2) Effective-region clipping}
		
		\FORALL{dihedrals and trihedrals $s_m\in\mathscr{S}$}
		\FORALL{constituent plane pairs $(s_1,s_2)\in s_m$}
		\STATE Compute intersection segment $[\mathbf{p}_{s1},\mathbf{p}_{s2}]$. (Eq.~\eqref{eq:intersection_segment})
		\FORALL{plane $s\in\{s_1,s_2\}$}
		\STATE Update clipped primitive $s\leftarrow s'$. (Eqs.~\eqref{eq:uv_frame}--~\eqref{eq:update_c})
		\ENDFOR
		\ENDFOR
		\ENDFOR
		
		\vspace{0.3em}
		\STATE \textbf{(3) Scatterer parameter alignment}
		
		\FORALL{multiple-bounce scatterers $s_m\in\mathscr{S}$}
		\STATE Align parameters to the canonical scattering center models.
		\ENDFOR
		
		\vspace{0.3em}
		\RETURN $\mathscr{S}$
		
	\end{algorithmic}
\end{algorithm}

\subsection{Effective-Region Clipping}
\begin{figure}[t]
	\centering
	\includegraphics[width=0.9\linewidth]{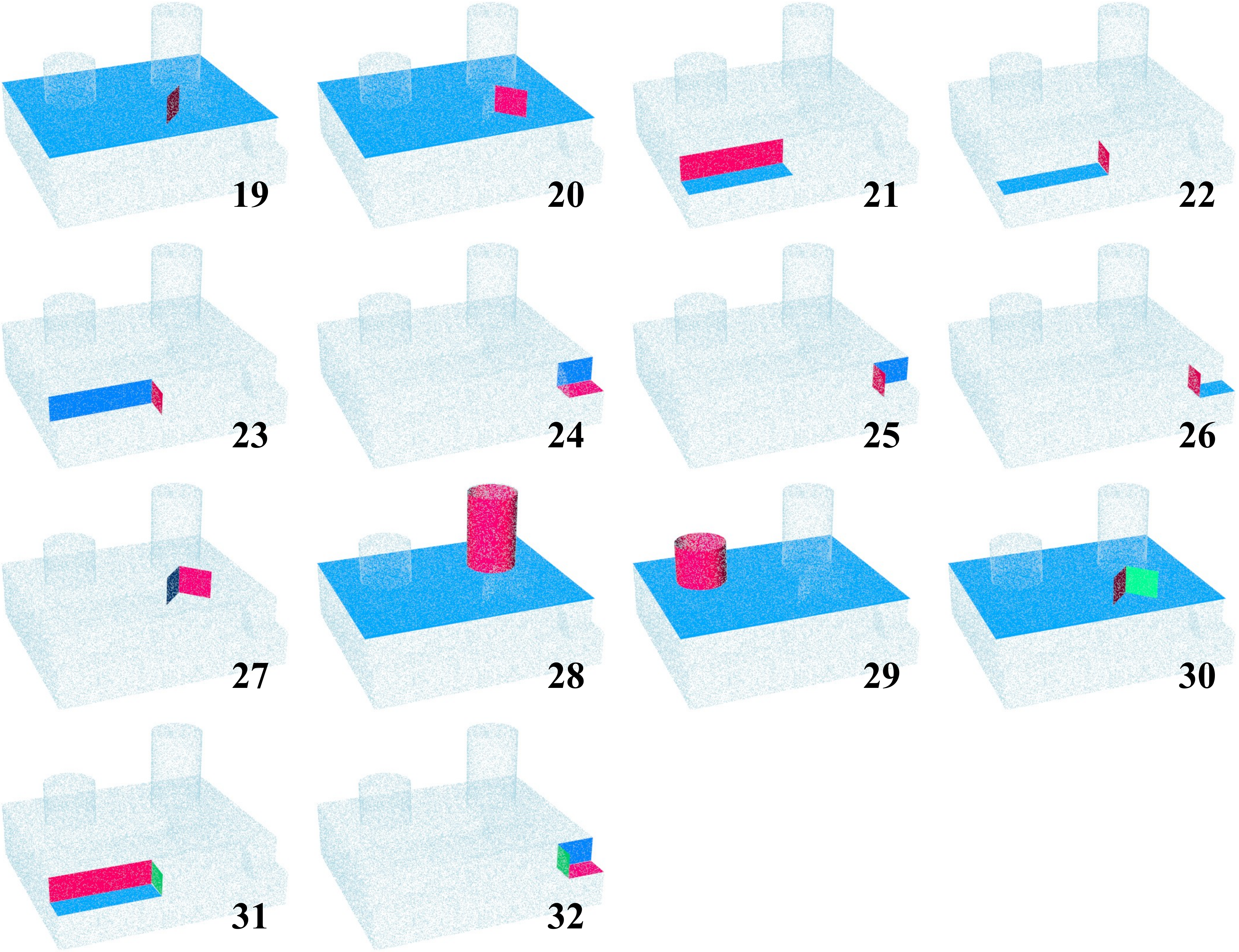}
	\caption{Potential multiple-bounce scatterers detected on the SLICY model: 9 dihedrals, 2 tophats, and 3 trihedrals. Indices 0--18 denote single-bounce scatterers, while indices from 19 denote multiple-bounce scatterers.}
	\label{fig:slicy_multiples}
\end{figure}

The multiple-bounce scatterers identified through geometric coupling do not yet constitute complete entities suitable for direct scattering computation.
For example, the dihedral labeled 20 in Fig.~\ref{fig:slicy_multiples} cannot be accurately represented by the two raw planar primitives alone. To address this issue, we introduce an effective-region clipping strategy that extracts only the portions actively participating in multiple-scattering.

\subsubsection{Intersection Segment Computation}
\begin{figure}[t]
	\centering
	\includegraphics[width=\linewidth]{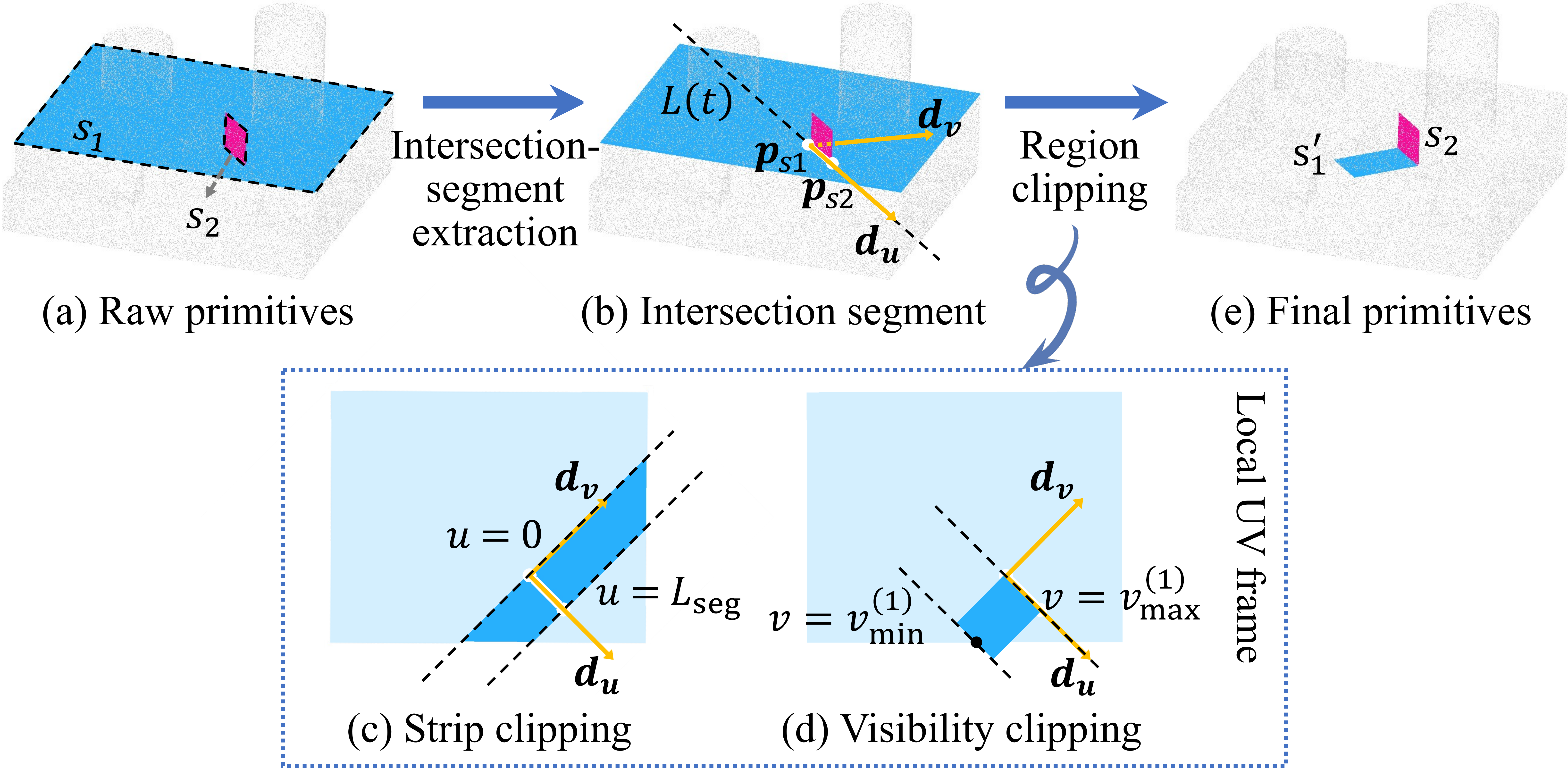}
	\caption{Effective-region clipping for a dihedral (illustrated using plane $s_1$). (a) Raw primitives. (b) Intersection line and local $(\mathbf{d}_u,\mathbf{d}_v)$ frame. (c) Clipping along the $u$-direction. (d) Clipping along the $v$-direction. (e) Back-projected effective region $s_1'$.}
	\label{fig:dihedral_clipping}
\end{figure}

Consider a dihedral formed by two planes $s_1=\mathrm{Plane}(\mathbf{c}_1,\mathbf{a}_1,\mathbf{d}_{11},\allowbreak
\mathbf{d}_{12},l_{11},l_{12})$ and $s_2=\mathrm{Plane}(\mathbf{c}_2,\mathbf{a}_2,\allowbreak
\mathbf{d}_{21},\mathbf{d}_{22},l_{21},l_{22})$ as illustrated in Fig.~\ref{fig:dihedral_clipping}(a). Their intersection
line is written as
\begin{equation}
	\mathbf{L}(t)
	=
	\mathbf{p}_{\mathrm{line}}
	+t\mathbf{d}_{\mathrm{line}},
	\qquad
	\mathbf{d}_{\mathrm{line}}
	=
	\frac{\mathbf{a}_1\times\mathbf{a}_2}
	{\|\mathbf{a}_1\times\mathbf{a}_2\|},
\end{equation}
where $\mathbf{p}_{\mathrm{line}}$ is obtained by solving the two plane
equations.

Intersecting $\mathbf{L}(t)$ with the rectangular boundaries of $s_1$ and $s_2$ gives the intervals $[t_{\min}^{(1)},t_{\max}^{(1)}]$ and $[t_{\min}^{(2)},t_{\max}^{(2)}]$, respectively. Their overlap defines the
shared intersection segment:
\begin{equation}
	\begin{aligned}
		\mathbf{p}_{s1} &= \mathbf{p}_{\text{line}}
		+ \max\!\left(t_{\min}^{(1)},t_{\min}^{(2)}\right)\mathbf{d}_{\text{line}},\\
		\mathbf{p}_{s2} &= \mathbf{p}_{\text{line}}
		+ \min\!\left(t_{\max}^{(1)},t_{\max}^{(2)}\right)\mathbf{d}_{\text{line}},
	\end{aligned}
	\label{eq:intersection_segment}
\end{equation}
whose length is $L_{\mathrm{seg}}=\|\mathbf{p}_{s2}-\mathbf{p}_{s1}\|$.

The shared segment determines the common extent of the two planes along their intersection direction. Each constituent plane is then clipped in two stages: it is first restricted to this common extent and subsequently restricted to the half-plane facing the other primitive.

\subsubsection{Strip Clipping}
Taking plane $s_1$ as an example, a local orthonormal frame is constructed as
\begin{equation}
	\mathbf{d}_u=\frac{\mathbf{p}_{s2}-\mathbf{p}_{s1}}{\|\mathbf{p}_{s2}-\mathbf{p}_{s1}\|},\quad
	\mathbf{d}_v=\frac{\mathbf{a}_1\times\mathbf{d}_u}{\|\mathbf{a}_1\times\mathbf{d}_u\|}.
	\label{eq:uv_frame}
\end{equation}
The four vertices of $s_1$ are first projected onto the $(u,v)$ frame. A Sutherland–Hodgman–type polygon clipping algorithm is then used to compute the intersection between the projected polygon and the strip $0\le u\le L_{\mathrm{seg}}$, yielding the region of $s_1$ located within the strip (Fig.~\ref{fig:dihedral_clipping}(c)).

\subsubsection{Half-Plane Clipping}
After strip clipping, the feasible $v$-range may vary along $u$. To obtain a valid rectangular region, we sort the $u$-coordinates of the polygon vertices to define a set of consecutive $u$-intervals, evaluate the cross-sectional range $[v_{\min}(u_0),v_{\max}(u_0)]$ at the midpoint $u_0$ of each interval, and take the intersection of all such ranges. As illustrated in Fig.~\ref{fig:dihedral_clipping}(d), the half-plane facing $s_2$ is finally retained:
\begin{equation}
	\left(v_{\min}^{(1)},v_{\max}^{(1)}\right)
	=
	\begin{cases}
		(\max_{u_0}v_{\min}(u_0),\,0),
		& \mathbf{a}_2^\top\mathbf{d}_v<0,\\
		(0,\,\min_{u_0}v_{\max}(u_0)),
		& \mathbf{a}_2^\top\mathbf{d}_v\geq0.
	\end{cases}
	\label{eq:facing_side_bounds}
\end{equation}
The side length and center are then obtained as
\begin{equation}
	\label{eq:update_l}
	l' = v_{\max}^{(1)}-v_{\min}^{(1)},
\end{equation}
\begin{equation}
	\label{eq:update_c}
	\mathbf{c}_1'=\mathbf{p}_{s1}+\frac{L_{\mathrm{seg}}}{2}\mathbf{d}_u
	+\frac{v_{\max}^{(1)}+v_{\min}^{(1)}}{2}\mathbf{d}_v .
\end{equation}
The clipped primitive is thus updated as
$s_1'=\mathrm{Plane}(\mathbf{c}_1',\mathbf{a}_1,\allowbreak\mathbf{d}_u,\mathbf{d}_v,L_{\text{seg}},l')$.
The same procedure is applied to $s_2$.

For top-hat scatterers, no clipping is required. For trihedrals, the above clipping process is applied independently to each of the three constituent dihedrals.

\subsection{Scatterer Parameter Alignment}
After clipping, each multiple-scattering is described by a physically accurate geometric specification. However, canonical scattering center models~\cite{Jackson2010Canonical3DBistatic} impose parameter-sharing assumptions, as illustrated in Fig.~\ref{fig:multi_scatterers}. The parameters are therefore aligned with the corresponding canonical definitions.

For a dihedral, the scatterer position is set to the midpoint of the intersection segment, $l$ is assigned its length, and $h$ is chosen as the smaller plane height. For a trihedral, the position is defined by the intersection of three planes, while $h$ is set to the minimum edge length. For a top-hat, the position is defined by the intersection of the cylinder axis with the plane, and the parameters $h$ and $r$ are inherited from the cylinder.

At this stage, the modeling of potential target scattering structures is completed. It should be noted that the primitives and their combinations in the scatterer set $\mathscr{S}$ are uniquely derived from the target geometry and do not necessarily correspond to strong scattering centers under a fixed observation configuration; their actual scattering contributions are jointly determined by the observation setup and the scattering models. Consequently, $\mathscr{S}$ does not need to be reconstructed when the observation conditions change.

\subsection{Global Scattering Computation}
\label{sec:global_scattering}

For a global observation configuration $\Lambda^g$, the incident and scattered directions are transformed into the local frame of each scatterer $s_m$ using the rotation matrix $\mathbf{R}_m$ determined by its geometric orientation. The resulting local angular configuration $\Lambda_m=(\theta_{i,m},\theta_{s,m},\phi_{i,m},\phi_{s,m})$ is then substituted into the corresponding canonical model. The total target response is obtained by coherent superposition:
\begin{equation}
	S_{\mathrm{target}}(f,\Lambda^g)
	=
	\sum_{s_m\in\mathscr{S}}
	S(f,\Lambda_m;\boldsymbol{\Theta}_m),
	\label{eq:total_response}
\end{equation}
where $S(\cdot)$ is given by~\eqref{eq:canonical_scatter}.

It should be noted that the framework evaluates each scatterer independently without explicitly accounting for global occlusion by other structures, which may overestimate some potential scatterers' contributions under certain views.


\section{Validations}
\label{sec:exp}

To validate the effectiveness of the proposed framework, several test cases are investigated, including the MSTAR SLICY model, a modified SLICY model, and simplified ship and aircraft targets. The CAD models of SLICY, modified SLICY, and the ship are manually constructed and sampled into point clouds, while the aircraft model is directly adopted from ShapeNet~\cite{Chang2015ShapeNet}. 
For each target, scatterer parameters are generated and used to compute SAR images under different observation configurations.

For the accuracy of scatterer geometric parameters, since the CAD ground-truth is available for SLICY, Modified SLICY, and the ship model, three types of error metrics are defined, corresponding to position, orientation, and size parameters:
\begin{subequations}
	\begin{align}
		e_c &= \|\mathbf{c}^{\mathrm{pred}} - \mathbf{c}^{\mathrm{gt}}\|_2, 
		\label{eq:pos_error}\\
		e_a &= \arccos\!\big(|\mathbf{a}^{\mathrm{pred}}\cdot\mathbf{a}^{\mathrm{gt}}|\big), 
		\label{eq:ori_error}\\
		e_l &= |l^{\mathrm{pred}} - l^{\mathrm{gt}}|.
		\label{eq:size_error}
	\end{align}
\end{subequations}

For the accuracy of scattering responses, all simulated results are compared with either RL-GO simulations in FEKO or measured data, depending on the experiment.
For quantitative evaluation, the normalized cross-correlation coefficient is adopted as the image similarity metric:
\begin{equation}
	\mathrm{Cor} =
	\frac{\sum_{x,y}
		\left(I(x,y)-\mu_I\right)
		\left(I^{\mathrm{ref}}(x,y)-\mu_{I^{\mathrm{ref}}}\right)}
	{\sqrt{
			\sum_{x,y}\left(I(x,y)-\mu_I\right)^2
			\sum_{x,y}\left(I^{\mathrm{ref}}(x,y)-\mu_{I^{\mathrm{ref}}}\right)^2}},
	\label{eq:correlation}
\end{equation}
where $I(x,y)$ denotes the SAR image generated by the proposed method, $I^{\mathrm{ref}}(x,y)$ denotes FEKO-simulated or MSTAR-measured reference image, and $\mu_I$ and $\mu_{I^{\mathrm{ref}}}$ represent the mean values of the two images, respectively.
To further evaluate the scattering-center distribution, we introduce peak matching metrics. Local peaks with amplitudes no lower than $-20$ dB relative to the maximum are extracted from both SAR images and matched by Euclidean distance. The matching recall, precision, and average localization error are defined as
\begin{subequations}
	\begin{gather}
		R_{\mathrm{m}} = \frac{N_\mathrm{m}}{N_{\mathrm{ref}}},
		\label{eq:match_recall}\\
		P_{\mathrm{m}} = \frac{N_\mathrm{m}}{N_{\mathrm{pred}}},
		\label{eq:match_precision}\\
		E_{\mathrm{loc}} = \frac{1}{N_\mathrm{m}}\sum_{(i,j)\in\mathcal{M}}
		\left\|\mathbf{p}^{\mathrm{pred}}_i-\mathbf{p}^{\mathrm{ref}}_j\right\|_2
		\label{eq:loc_error}
	\end{gather}
\end{subequations}
where $N_\mathrm{m}$ is the number of matched peaks, $N_{\mathrm{ref}}$ and $N_{\mathrm{pred}}$ are the numbers of peaks in the reference and predicted images, $\mathcal{M}$ is the set of matched peak pairs, and $\mathbf{p}^{\mathrm{pred}}_i$ and $\mathbf{p}^{\mathrm{ref}}_j$ are their pixel coordinates.

The RANSAC parameters are empirically set as follows.
Since primitive fitting relies on distance consistency, normal consistency, and connectivity, the input point cloud should cover the main target surfaces and provide stable normals. Uniform or extremely dense sampling is not required, but each primitive of interest should contain more points than the minimum support threshold $\tau$; otherwise, small local structures may be missed.
Accordingly, the distance threshold $\varepsilon$ is set to $0.001$ times the target scale, and the normal consistency threshold $\alpha$ is set to $0.99$. When higher geometric accuracy or finer structural separation is required, a smaller $\varepsilon$ and a larger $\alpha$ can be adopted.
The connectivity bitmap resolution $\beta$ is set to $0.01$ times the target scale. For sparse point clouds, a larger $\beta$ is used to avoid fragmenting a single primitive. The confidence threshold $\eta$ is set to $0.95$. The minimum support $\tau$ is set to $100$ and adjusted according to the point cloud size.

Most test cases follow these settings. For the modified SLICY model, $\beta$ is set to $0.05$. For the aircraft target, $\tau$ is reduced to $25$ to preserve small scattering-related structures in the engine regions, and $\beta$ is set to $0.003$ to separate physically disconnected structures lying on similar planar equations.

\subsection{SLICY}

\begin{figure}[t]
	\centering
	\includegraphics[width=0.45\linewidth]{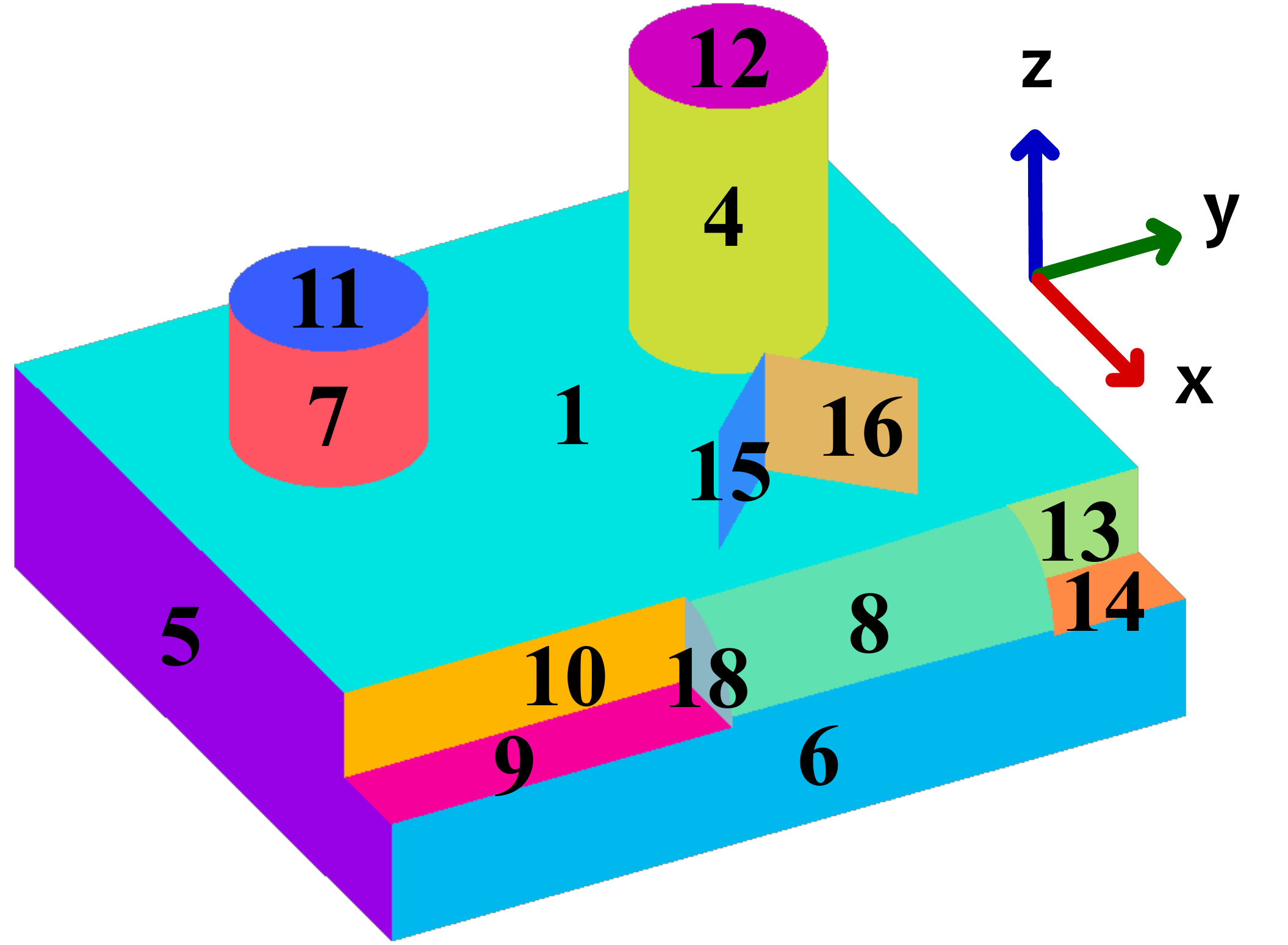}
	\caption{Primitive fitting result of the SLICY model. The 19 detected primitives are shown in different colors, with visible primitive indices labeled.}
	\label{fig:slicy_prims}
\end{figure}

\begin{table}[t]
	\centering
	\caption{Comparison between estimated and ground-truth parameters of selected primitives in the SLICY model.}
	\label{tab:slicy_param_error}
	\scriptsize
	\setlength{\tabcolsep}{1.2pt}
	\renewcommand{\arraystretch}{1.1}
	\begin{tabular}{@{}c c c c c c}
		\hline
		Index & Type & Param. & Estimated value & Ground truth & Error \\
		\hline
		
		\multirow{6}{*}{0} & \multirow{6}{*}{Plane} 
		& $\mathbf{c}$   & $(2.4999,\,2.8123,\,0)$ & $(2.5,\,2.8125,\,0)$ & 0.0002 \\
		& & $\mathbf{a}$   & $(0,\,0,\,-1)$ & $(0,\,0,\,-1)$ & 0 \\
		& & $\mathbf{d}_1$ & $(1,\,0,\,0)$ & $(1,\,0,\,0)$ & 0 \\
		& & $\mathbf{d}_2$ & $(0,\,1,\,0)$ & $(0,\,1,\,0)$ & 0 \\
		& & $l_1$ & 4.9998 & 5 & 0.0002 \\
		& & $l_2$ & 5.624 & 5.625 & 0.001 \\
		\hline
		
		\multirow{6}{*}{1} & \multirow{6}{*}{Plane} 
		& $\mathbf{c}$   & $(2.2382,\,2.8123,\,1.4859)$ & $(2.1885,\,2.8125,\,1.486)$ & 0.0497 \\
		& & $\mathbf{a}$   & $(0.0001,\,0,\,1)$ & $(0,\,0,\,1)$ & 0 \\
		& & $\mathbf{d}_1$ & $(1,\,0.0002,\,0.0001)$ & $(1,\,0,\,0)$ & 0.0002 \\
		& & $\mathbf{d}_2$ & $(-0.0002,\,1,\,0)$ & $(0,\,1,\,0)$ & 0.0001 \\
		& & $l_1$ & 4.4750 & 4.377 & 0.098 \\
		& & $l_2$ & 5.6238 & 5.625 & 0.0012 \\
		\hline
		
		\multirow{4}{*}{4} & \multirow{4}{*}{Cylinder} 
		& $\mathbf{c}$ & $(1.6477,\,4.1776,\,2.4431)$ & $(1.65,\,4.1775,\,2.461)$ & 0.018 \\
		& & $\mathbf{a}$ & $(0,\,0,\,1)$ & $(0,\,0,\,1)$ & 0 \\
		& & $r$ & 0.6238 & 0.625 & 0.0012 \\
		& & $h$ & 1.9233 & 1.95 & 0.0267 \\
		\hline
		
		\multirow{4}{*}{7} & \multirow{4}{*}{Cylinder} 
		& $\mathbf{c}$ & $(1.648,\,1.35,\,1.9907)$ & $(1.65,\,1.3525,\,1.986)$ & 0.0057 \\
		& & $\mathbf{a}$ & $(0,\,0,\,1)$ & $(0,\,0,\,1)$ & 0 \\
		& & $r$ & 0.6238 & 0.625 & 0.0012 \\
		& & $h$ & 0.9941 & 1 & 0.0059 \\
		\hline
		
		\multirow{4}{*}{8} & \multirow{4}{*}{Cylinder} 
		& $\mathbf{c}$ & $(4.38,\,3.5182,\,0.8648)$ & $(4.377,\,3.5525,\,0.863)$ & 0.0345 \\
		& & $\mathbf{a}$ & $(-0.001,\,1.0,\,-0.001)$ & $(0,\,1,\,0)$ & 0.0014 \\
		& & $r$ & 0.6193 & 0.623 & 0.0037 \\
		& & $h$ & 2.2627 & 2.279 & 0.0163 \\
		\hline
		
	\end{tabular}
\end{table}

The first example is the SLICY model from the MSTAR database~\cite{Diemunsch1998MSTARATR}. The target size is $5\,\mathrm{m}\times5.625\,\mathrm{m}\times3.436\,\mathrm{m}$. It contains major scattering structures in canonical scattering center models, including planes, cylinders, dihedrals, trihedrals, and top-hats, and is therefore widely used as a standard benchmark in scattering modeling studies.

\begin{figure}[t]
	\centering
	\includegraphics[width=0.85\linewidth]{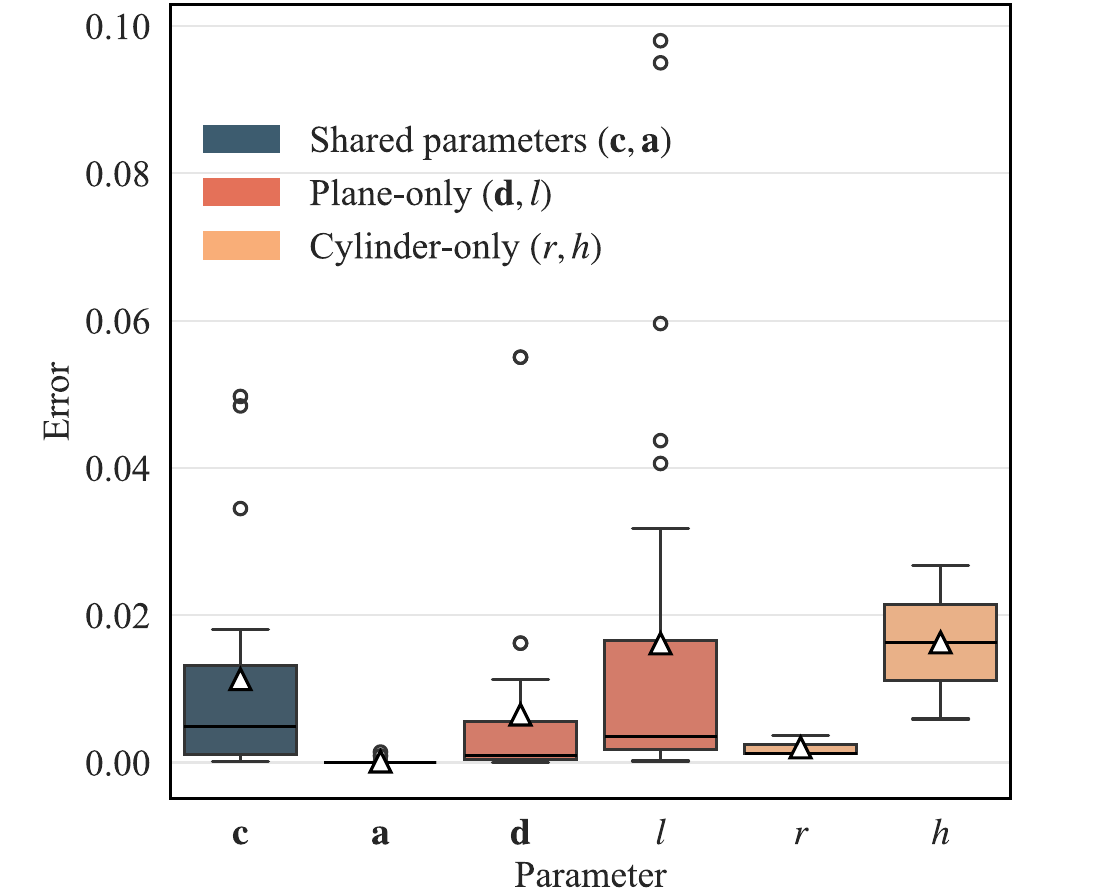}
	\caption{Box plots of primitive parameter fitting errors on the SLICY model, including center positions \(\mathbf{c}\) and orientations \(\mathbf{a}\) for all 19 primitives, side directions \(\mathbf{d}\) and side lengths \(l\) for 16 planes, and radii \(r\) and heights \(h\) for 3 cylinders.}
	\label{fig:slicy_boxplot}
\end{figure}

\begin{table}[t!]
	\centering
	\caption{Parameters of scatterers in the SLICY model.}
	\label{tab:slicy_scatterers}
	\scriptsize
	\setlength{\tabcolsep}{5.6pt}
	\renewcommand{\arraystretch}{1.12}
	\begin{tabular}{c c c c c c}
		\hline
		Index & Type & Position (m) & $L$ (m) & $H$ (m) & $r$ (m) \\
		\hline
		0  & Plane     & $(2.4999,\,2.8123,\,0)$        & 4.9998 & 5.624  & -- \\
		1  & Plane     & $(2.2382,\,2.8123,\,1.4859)$   & 4.475  & 5.6238 & -- \\
		2  & Plane     & $(0,\,2.8116,\,0.7422)$        & 5.6229 & 1.4841 & -- \\
		3  & Plane     & $(2.4998,\,5.625,\,0.7434)$    & 4.9966 & 1.4844 & -- \\
		4  & Cylinder  & $(1.6477,\,4.1776,\,2.4431)$   & --     & 1.9233 & 0.6238 \\
		5  & Plane     & $(2.5,\,0,\,0.7431)$            & 4.9987 & 1.4841 & -- \\
		6  & Plane     & $(4.9998,\,2.8115,\,0.4799)$   & 5.6223 & 0.958  & -- \\
		7  & Cylinder  & $(1.648,\,1.35,\,1.9907)$       & --     & 0.9941 & 0.6238 \\
		8  & Cylinder  & $(4.38,\,3.5182,\,0.8648)$      & --     & 2.2627 & 0.6193 \\
		9  & Plane     & $(4.6887,\,1.2058,\,0.863)$     & 2.4065 & 0.6217 & -- \\
		10 & Plane     & $(4.377,\,1.2081,\,1.1725)$     & 2.4094 & 0.62   & -- \\
		11 & Plane     & $(1.6531,\,1.3562,\,2.486)$     & 1.2357 & 1.2063 & -- \\
		12 & Plane     & $(1.64,\,4.1764,\,3.436)$       & 1.2255 & 1.2094 & -- \\
		13 & Plane     & $(4.377,\,5.1607,\,1.1737)$     & 0.9271 & 0.6138 & -- \\
		14 & Plane     & $(4.6875,\,5.1582,\,0.863)$     & 0.9314 & 0.6212 & -- \\
		15 & Plane     & $(3.5794,\,3.2731,\,1.8364)$    & 0.695  & 0.7861 & -- \\
		16 & Plane     & $(3.5833,\,3.8358,\,1.8309)$    & 0.8005 & 0.6865 & -- \\
		17 & Plane     & $(4.6806,\,4.692,\,1.1667)$     & 0.5912 & 0.5995 & -- \\
		18 & Plane     & $(4.6933,\,2.413,\,1.1601)$     & 0.6322 & 0.5634 & -- \\
		19 & Dihedral  & $(3.5822,\,3.2703,\,1.4858)$    & 0.7862 & 0.695  & -- \\
		20 & Dihedral  & $(3.5821,\,3.8347,\,1.4858)$    & 0.8005 & 0.6865 & -- \\
		21 & Dihedral  & $(4.3779,\,1.2058,\,0.863)$     & 2.4061 & 0.6195 & -- \\
		22 & Dihedral  & $(4.6852,\,2.409,\,0.863)$      & 0.6166 & 0.5625 & -- \\
		23 & Dihedral  & $(4.377,\,2.4128,\,1.1775)$     & 0.5643 & 0.6312 & -- \\
		24 & Dihedral  & $(4.377,\,5.1574,\,0.8668)$     & 0.9314 & 0.6138 & -- \\
		25 & Dihedral  & $(4.377,\,4.6972,\,1.1624)$     & 0.5984 & 0.5911 & -- \\
		26 & Dihedral  & $(4.6855,\,4.6925,\,0.863)$     & 0.5913 & 0.5994 & -- \\
		27 & Dihedral  & $(3.3015,\,3.551,\,1.8328)$     & 0.6865 & 0.786  & -- \\
		28 & Top-hat   & $(1.6477,\,4.1776,\,1.486)$     & --     & 1.9233 & 0.6238 \\
		29 & Top-hat   & $(1.648,\,1.35,\,1.486)$        & --     & 0.9941 & 0.6238 \\
		30 & Trihedral & $(3.3,\,3.5525,\,1.4858)$       & --     & 0.6865 & -- \\
		31 & Trihedral & $(4.377,\,2.413,\,0.863)$       & --     & 0.5634 & -- \\
		32 & Trihedral & $(4.377,\,4.692,\,0.863)$       & --     & 0.5912 & -- \\
		\hline
	\end{tabular}
\end{table}

In the experiments, a point cloud with 50\,000 points is uniformly sampled from the SLICY model. Fig.~\ref{fig:slicy_prims} shows the primitive fitting results, where 16 planes and 3 cylinders are detected, fully covering the geometric surfaces of the target.

To avoid enumerating all 19 primitives, Table~\ref{tab:slicy_param_error} reports five representative cases (the top and bottom base planes and three cylinders) together with the estimated parameters and CAD ground truth. 
Overall, the parameters are accurately recovered. 
The relatively larger errors of Plane~1 in the $x$-component of the center and in $l_1$ are mainly caused by support contamination: points near the upper rim of the adjacent horizontal cylinder have normals close to $+z$ and are therefore partially absorbed into the inlier set of Plane~1, leading to a slight overestimation of its extent along the $x$ direction.

Fig.~\ref{fig:slicy_boxplot} further summarizes the fitting errors over all primitives. 
The orientation estimates are the most stable, with an average error of \(1.0\times10^{-4}\,\mathrm{rad}\). 
The mean position error is \(0.0113\,\mathrm{m}\), with two outliers from Plane~1 and Plane~6 caused by absorption of adjacent cylindrical edge points. 
For the 16 planes, the mean errors of edge directions and side lengths are \(0.0065\,\mathrm{rad}\) and \(0.0162\,\mathrm{m}\), respectively; for the 3 cylinders, the mean radius and height errors are \(0.002\,\mathrm{m}\) and \(0.016\,\mathrm{m}\). 
Overall, the mean, median, and interquartile ranges of all parameter errors are below \(0.02\,\mathrm{m}\) or \(0.01\,\mathrm{rad}\) (\(\approx0.579^\circ\)), and even the outliers remain within \(0.1\,\mathrm{m}\) or \(0.02\,\mathrm{rad}\). 
These results confirm that the RANSAC-based detection and closed-form estimation can stably recover primitives and provide reliable geometric inputs for scatterer construction.

\begin{figure*}[t]
	\centering
	\includegraphics[width=0.85\linewidth]{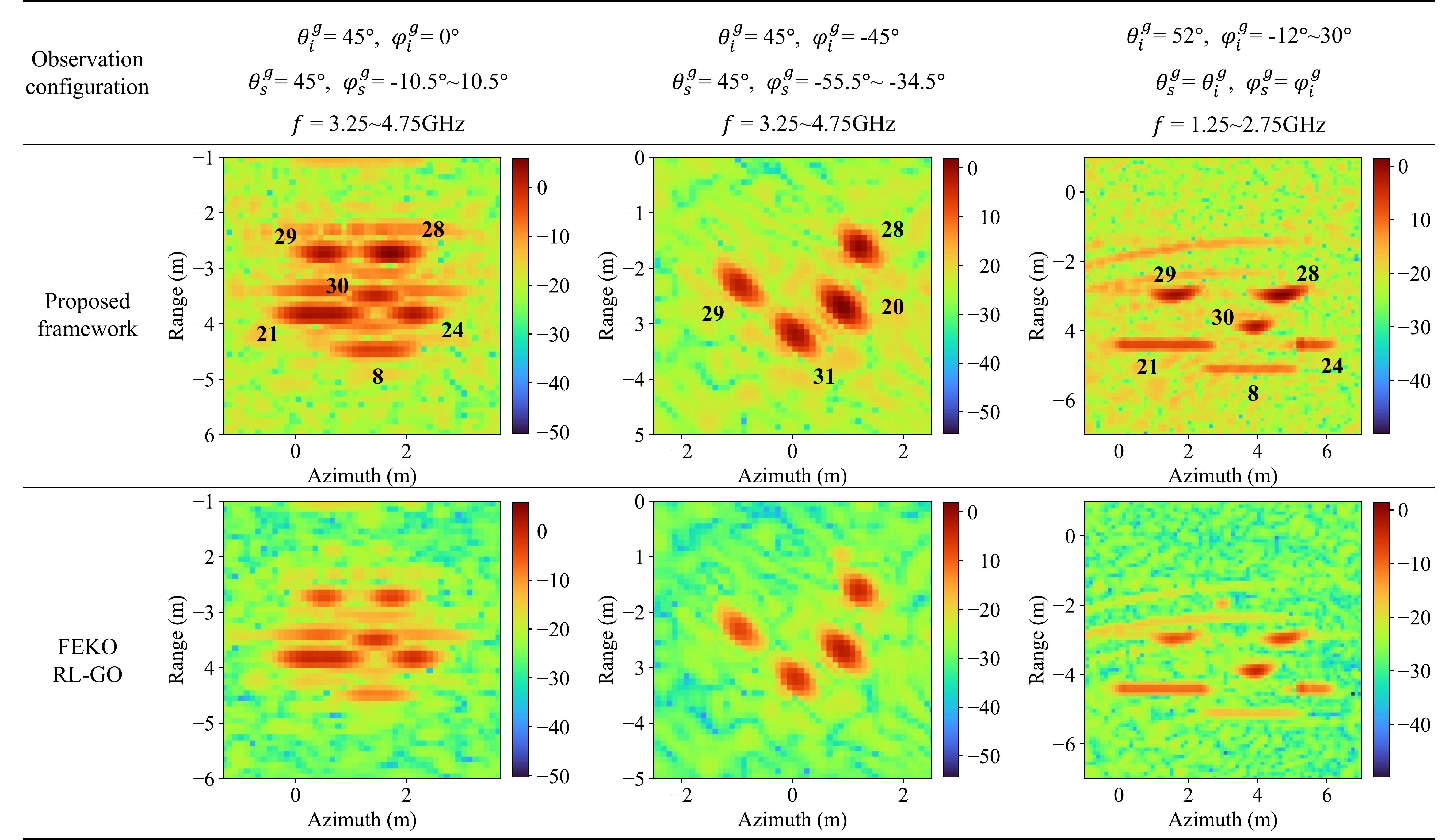}
	\caption{Simulated SAR images of the SLICY model under different observation configurations: settings (top), proposed method (middle), and FEKO (bottom).}
	\label{fig:slicy_sar}
\end{figure*}

Table~\ref{tab:slicy_scatterers} lists the complete set of scatterers uniquely determined for the SLICY model. 
This set includes all potential single- and multiple-bounce scattering structures and is fixed and independent of observation conditions. 

Fig.~\ref{fig:slicy_sar} presents the simulated SAR images of SLICY under three representative observation configurations, together with the FEKO results. 
In the first case (front–upper view), both the range and azimuth resolutions are $0.1$\,m. 
Six dominant scatterers are clearly observed, including two top-hats (indices 28 and 29) and one trihedral (index 30) appearing as localized scattering centers, and two dihedrals (indices 21 and 24) and one cylinder (index 8) appearing as distributed scattering centers. 
Other potential scatterers either face away from the incident direction (e.g., the bottom plane, index 0) or produce weak contributions under this observation (e.g., the two trihedrals formed by the base and the horizontal cylinder, indices 31 and 32), and therefore produce negligible contributions.  
The SAR image obtained by the proposed method agrees well with FEKO, with an image similarity of 0.8623.

The second column of Fig.~\ref{fig:slicy_sar} corresponds to a front-side observation configuration. 
Under this observation, four scatterers dominate the image, including two top-hats (indices 28 and 29), one dihedral (index 20), and one trihedral (index 31). 
The proposed method again exhibits high consistency with FEKO, yielding an image similarity of 0.9983.

The third column shows a monostatic imaging case. 
Although the spatial distributions and scattering patterns vary with the observation, the dominant contributors observed in the first case remain the primary scattering sources, and their relative strengths are largely preserved. 
The resulting image similarity with FEKO is 0.8544.

\subsection{Modified SLICY}

\begin{figure}[t]
	\centering
	\includegraphics[width=0.4\linewidth]{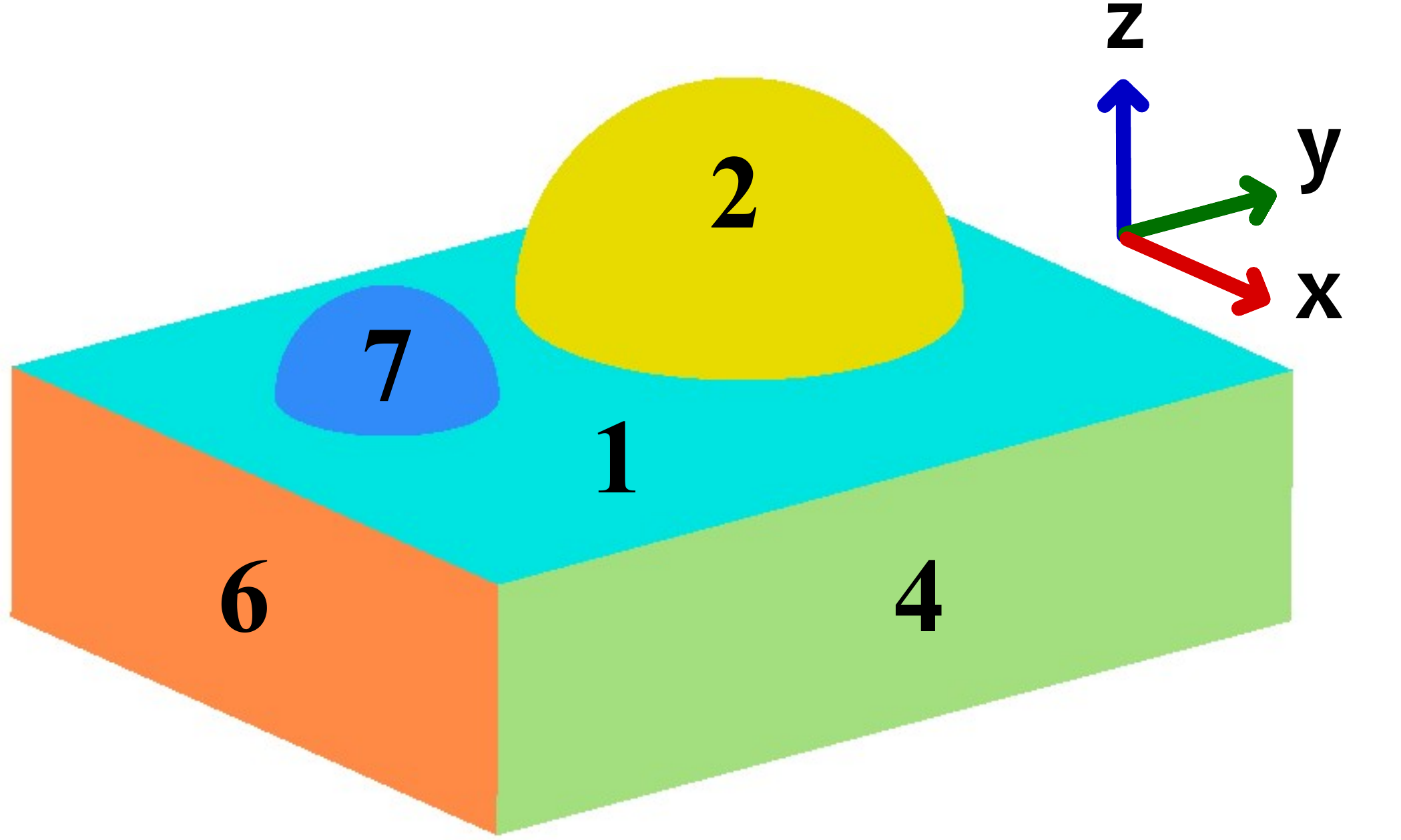}
	\caption{Primitive fitting result of the modified SLICY model. The 8 detected primitives are shown in different colors, with visible primitive indices labeled.}
	\label{fig:modi_slicy_prims}
\end{figure}

The modified SLICY model, with an overall size of \(4.377\,\mathrm{m}\times5.625\,\mathrm{m}\times2.736\,\mathrm{m}\), is used to evaluate spherical primitive extraction and weak-scattering modeling. 
A point cloud of 10,000 points is sampled, from which six planes and two spheres are correctly detected, as shown in Fig.~\ref{fig:modi_slicy_prims}. Table~\ref{tab:modi_slicy_param_error} reports the estimated parameters of the two spheres, with both the position and radius errors within \(0.005\,\mathrm{m}\). 
No multiple-bounce scatterers are detected, as expected. 

Under the same frontal-top observation setting as in the first column of Fig.~\ref{fig:slicy_sar}, two distributed scattering responses from the front face (index 4) and the top face (index 1) of the base, and two localized scattering responses from the two spheres (indices 2 and 7), can be clearly observed. The image similarity with FEKO is 0.8347, confirming the accurate modeling of spherical primitives and their scattering responses.

\begin{table}[t]
	\centering
	\caption{Comparison between estimated and ground-truth parameters of two spheres in the modified SLICY model.}
	\label{tab:modi_slicy_param_error}
	\scriptsize
	\setlength{\tabcolsep}{1.5pt}
	\renewcommand{\arraystretch}{1.15}
	\begin{tabular}{@{}c c c c c c}
		\hline
		Index & Type & Param. & Estimated value & Ground truth & Error \\
		\hline
		
		\multirow{2}{*}{2} & \multirow{2}{*}{Sphere} 
		& $\mathbf{c}$ & $(1.6485,\,3.8507,\,1.4862)$ & $(1.65,\,3.85,\,1.486)$ & 0.0016 \\
		& & $r$ & 1.249 & 1.25 & 0.001 \\
		\hline
		
		\multirow{2}{*}{7} & \multirow{2}{*}{Sphere} 
		& $\mathbf{c}$ & $(1.6502,\,1.3573,\,1.4849)$ & $(1.65,\,1.3525,\,1.486)$ & 0.0049 \\
		& & $r$ & 0.6249 & 0.625 & 0.0001 \\
		\hline
		
	\end{tabular}
\end{table}

\begin{figure}[t!]
	\centering
	\includegraphics[width=0.9\linewidth]{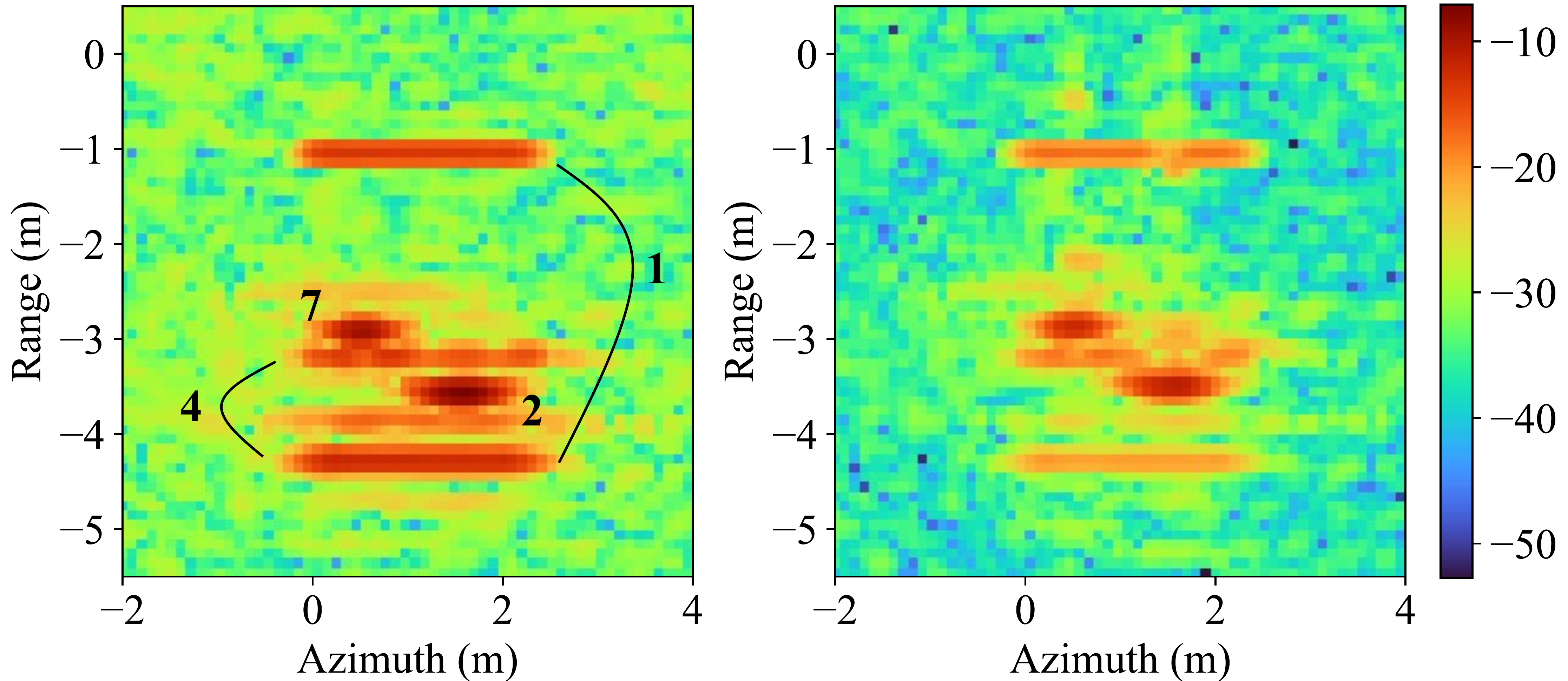}
	\caption{SAR simulation results of the modified SLICY model: proposed method (left), and FEKO (right).}
	\label{fig:modi_slicy_sar_img}
\end{figure}

\subsection{Ship}

The third example considers a simplified ship target, whose overall size is $12.6~\mathrm{m} \times 115.97~\mathrm{m} \times 24.13~\mathrm{m}$. The target primarily consists of planar and dihedral scattering structures and contains a larger number of primitives and much denser scattering structures compared with SLICY.

\begin{figure}[t]
	\centering
	\includegraphics[width=0.75\linewidth]{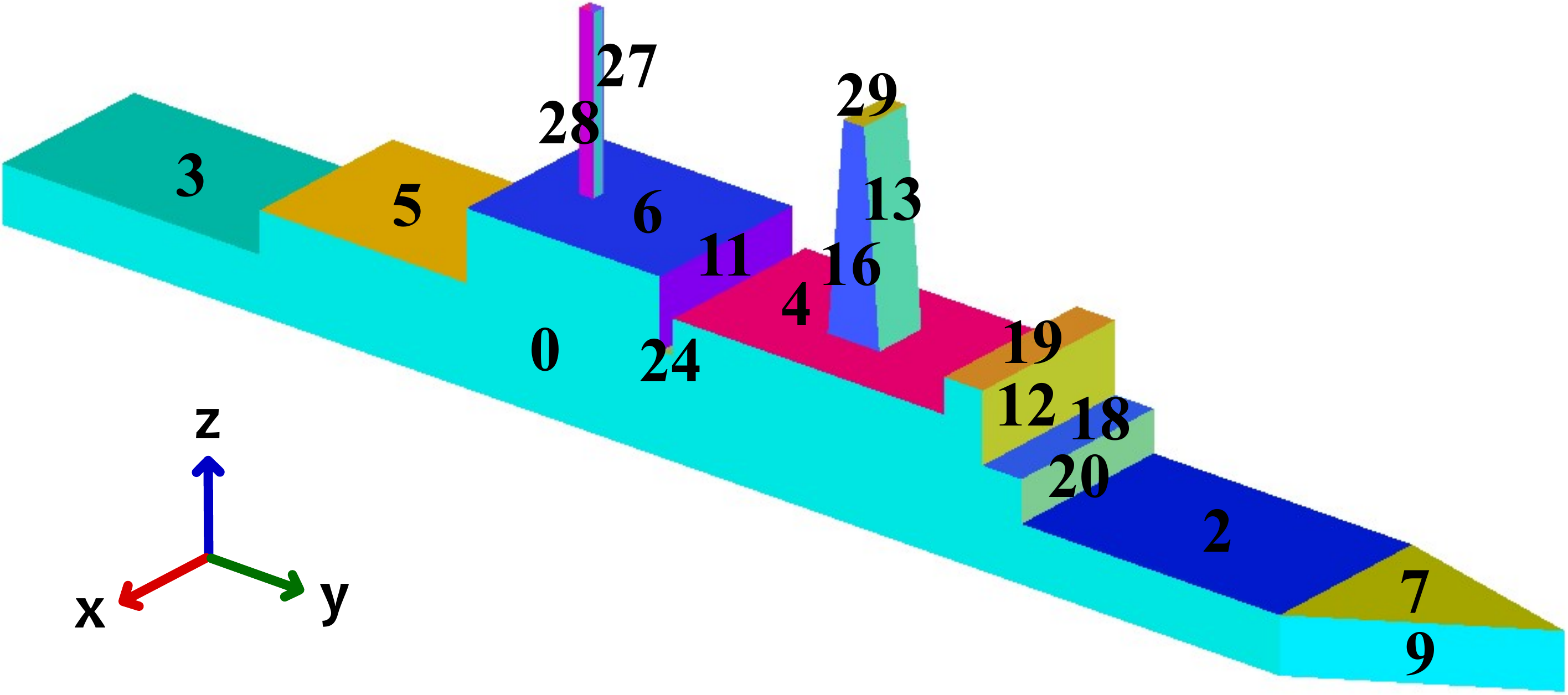}
	\caption{Primitive fitting result of the ship target. The 30 detected planar primitives are shown in different colors, with visible primitive indices labeled.}
	\label{fig:ship_prims}
\end{figure}

A point cloud of 50\,000 points is sampled. 
After primitive detection, the ship model is decomposed into 30 planes. 
Fig.~\ref{fig:ship_prims} shows the fitted planar surfaces after computing their intersection boundaries and clipping redundant regions.

Using~\eqref{eq:pos_error}–\eqref{eq:size_error}, the geometric estimation errors are computed for all planar primitives, and the average errors are summarized in Table~\ref{tab:ship_errors}. 
The orientation estimates remain highly stable. 
Despite the significantly larger scale of the ship compared with SLICY, the obtained position and size errors indicate that the proposed method can still accurately recover the geometric building blocks of large engineering targets.

\begin{table}[t]
	\centering
	\caption{Estimation errors of all planar primitives in the ship target.}
	\label{tab:ship_errors}
	\scriptsize
	\setlength{\tabcolsep}{13.1pt}
	\renewcommand{\arraystretch}{1.15}
	\begin{tabular}{c c c c c}
		\hline
		Parameter & $\mathbf{c}$ (m) & $\mathbf{a}$ (rad) & $\mathbf{d}$ (rad) & $l$ (m) \\
		\hline
		Mean error & 0.0321 & 0.0001 & 0.0068 & 0.0383 \\
		\hline
	\end{tabular}
\end{table}

\begin{figure}[t!]
	\centering
	\includegraphics[width=0.9\linewidth]{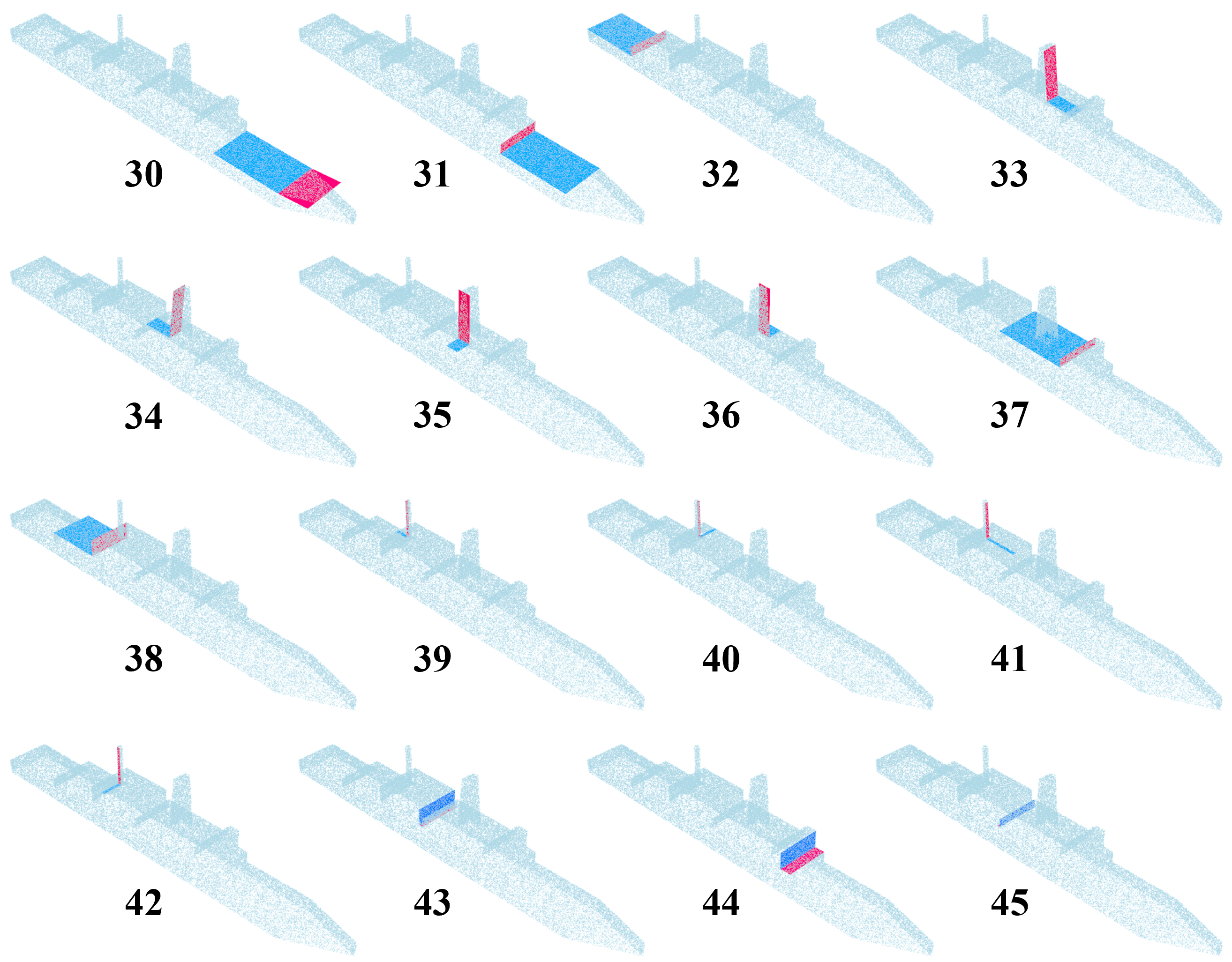}
	\caption{Potential multiple-bounce scatterers on the ship model: 16 dihedrals.}
	\label{fig:ship_mutliples}
\end{figure}

Based on the detected planar primitives, adjacency and directional constraints are applied to identify coupled primitive pairs. 
A total of 16 dihedrals are detected, as illustrated in Fig.~\ref{fig:ship_mutliples}. 
Combining single- and multiple-bounce structures, 46 potential scatterers are constructed, forming a unique scatterer set for the ship target.

\begin{figure*}[t]
	\centering
	\includegraphics[width=0.85\linewidth]{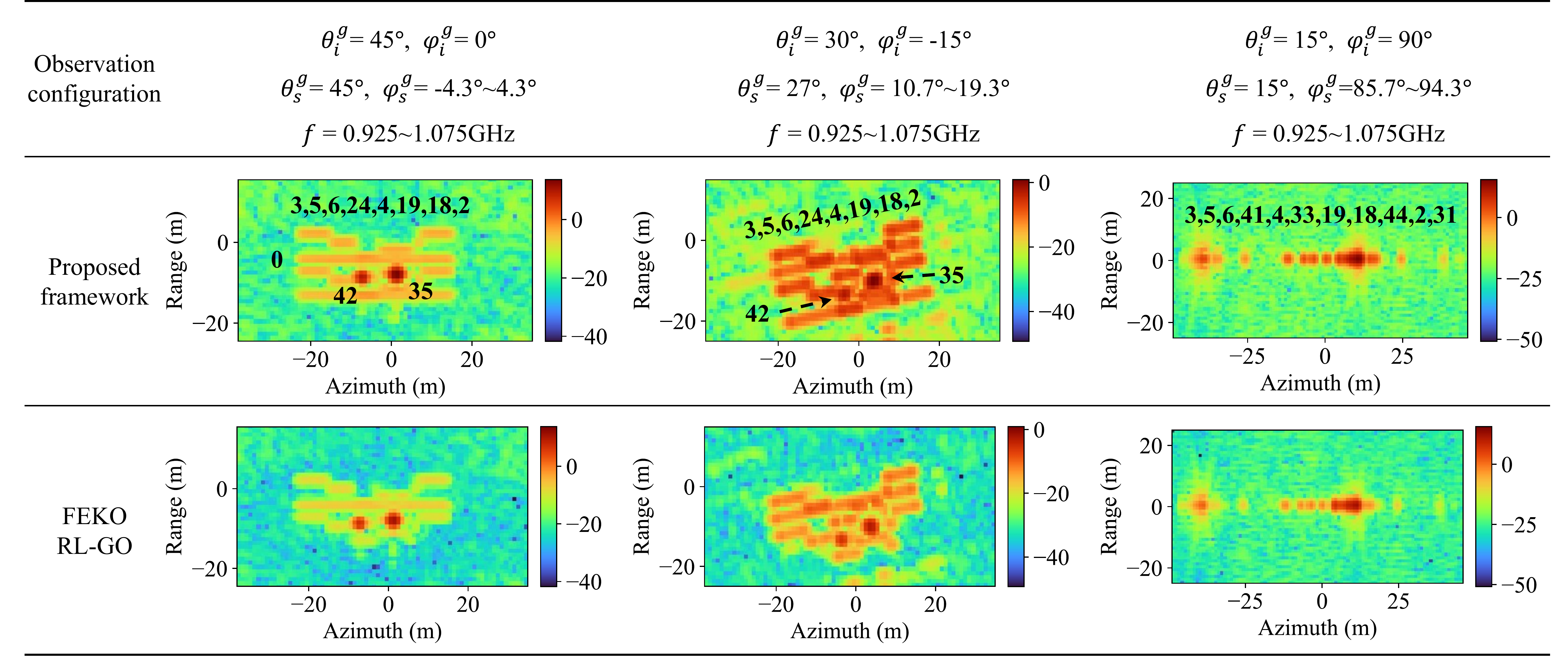}
	\caption{Simulated SAR images of the ship target under different observation configurations: settings (top), proposed method (middle), and FEKO (bottom).}
	\label{fig:ship_sar_img}
\end{figure*}

Fig.~\ref{fig:ship_sar_img} presents the simulated SAR images under three representative observation configurations, together with the reference results from FEKO. 
In the first configuration, both the range and azimuth resolutions are 1\,m. 
Two short dihedrals formed by the superstructure and the mast (indices 35 and 42) produce the strongest responses. 
In addition, the top surfaces of the deck and the superstructure (e.g., indices 3, 5, 6, 24, 4, 19, 18, and 2 from stern to bow), as well as the starboard side of the hull (e.g., indices 0, 16, and 28), also contribute noticeable scattering. 
These planar scatterers form clearly ordered distributed responses along the azimuth direction in the SAR image, exhibiting a strong spatial correspondence with the 3D geometry.

Compared with the FEKO results, an additional elongated distributed stripe appears at the bottom of the SAR image produced by the proposed method. 
This discrepancy arises from the geometric treatment of the irregular starboard side surface (index 0), which is approximated as its minimum-area bounding rectangle during primitive fitting. 
As a result, scattering contributions originally distributed along irregular boundaries are merged into a single continuous structure. 
The resulting image similarity is 0.9979.

Keeping the frequency settings and angular span unchanged, the proposed method is further evaluated under different incident and scattering angles. 
The dominant scatterers in the second column of Fig.~\ref{fig:ship_sar_img} remain largely consistent with those in the first case, mainly consisting of two dihedrals and several planar structures, with an image similarity of 0.9501. 
The third column corresponds to a frontal observation, where the dominant responses are produced by the upper planar surfaces of the deck and superstructure (e.g., indices 3, 5, 6, 4, 19, 18, and 2) together with several dihedral structures (e.g., indices 41, 33, 44, and 31). 
Under this configuration, the image similarity reaches 0.976. 

\begin{figure}[t]
	\centering
	\includegraphics[width=\linewidth]{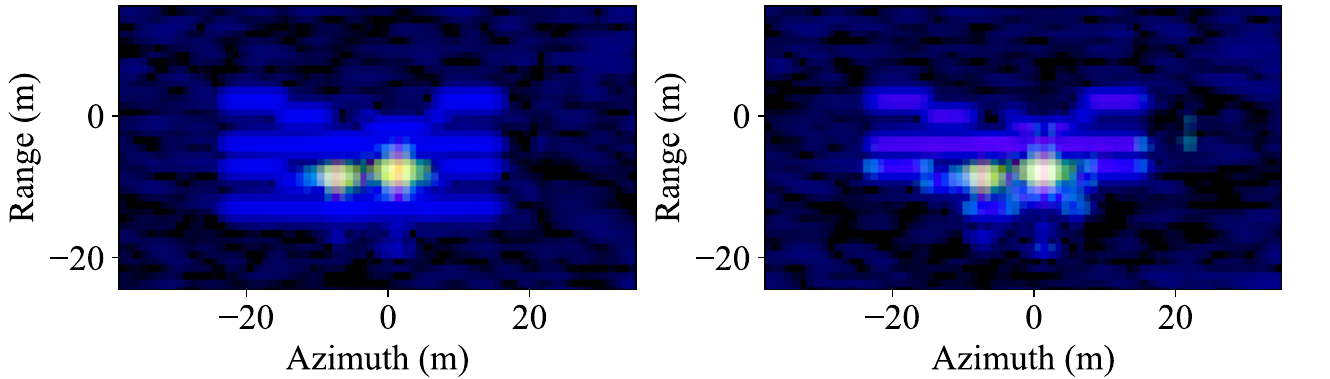}
	\caption{Comparison of full-polarimetric Pauli RGB pseudo-color images of the ship target generated by the proposed method (left) and FEKO (right).}
	\label{fig:ship_polsar}
\end{figure}

To evaluate full-polarimetric modeling, Fig.~\ref{fig:ship_polsar} compares the Pauli RGB pseudo-color images under the first observation setting in Fig.~\ref{fig:ship_sar_img}. The proposed framework and FEKO show consistent bright regions and color distributions, both reflecting the main planar single-bounce and mast-related dihedral responses. Minor local color differences mainly arise from weak scattering components or sidelobes across polarimetric channels.

\subsection{Aircraft}
\begin{figure}[t]
	\centering
	\includegraphics[width=\linewidth]{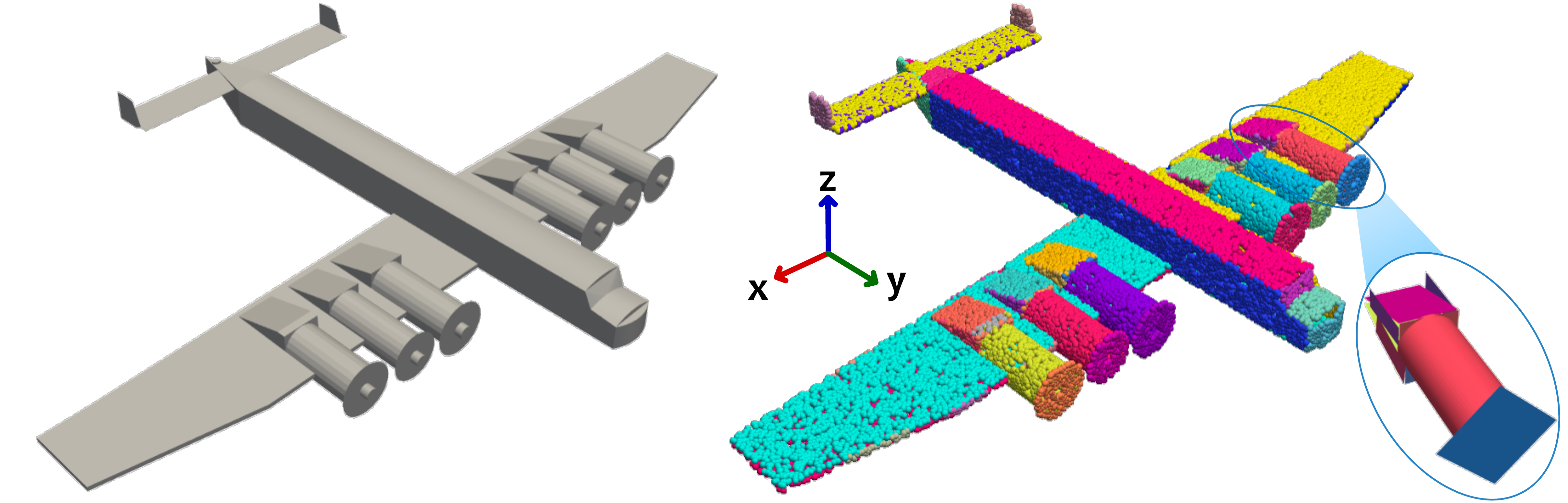}
	\caption{Original ShapeNet aircraft mesh (left) and primitive fitting result with 98 primitives (right). Different colors denote different parameterized primitives, with a zoomed-in view of one engine.}
	\label{fig:aircraft_prims}
\end{figure}

The fourth experiment considers an aircraft target (the Ju-390 bomber) for validation. As shown in Fig.~\ref{fig:aircraft_prims}, the mesh model is directly obtained from the ShapeNet dataset~\cite{Chang2015ShapeNet}, with overall dimensions of $49.07~\mathrm{m} \times 34.12~\mathrm{m} \times 5.27~\mathrm{m}$. 
Compared with SLICY and ship targets, the aircraft exhibits substantially higher geometric complexity, featuring both a larger number of primitives and a wide range of structural scales, from the overall fuselage and wings to fine components such as engines and their attachments.

\begin{figure*}[t]
	\centering
	\includegraphics[width=0.85\linewidth]{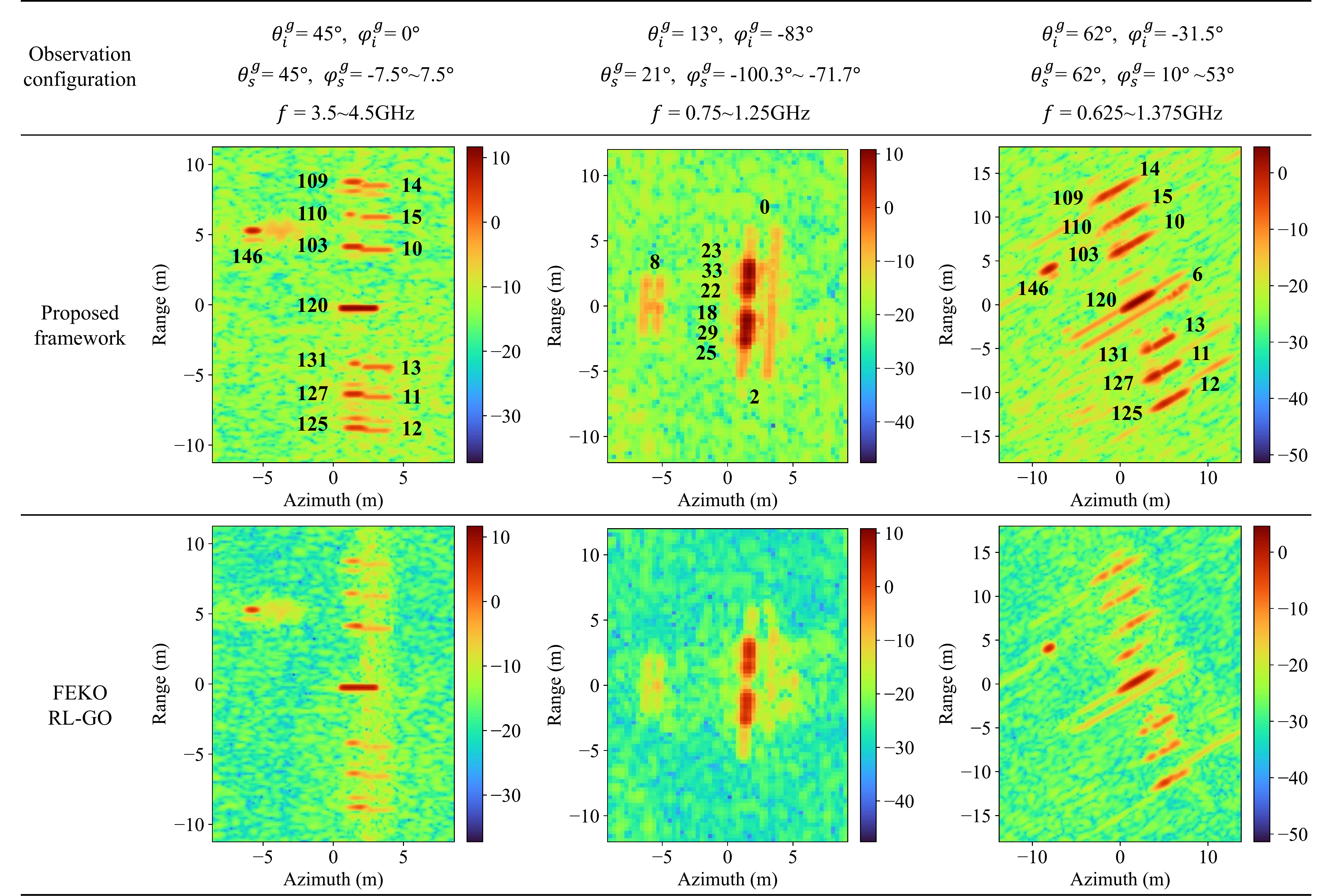}
	\caption{Simulated SAR images of the aircraft target under different observation configurations: settings (top), proposed method (middle), and FEKO (bottom).}
	\label{fig:aircraft_sar_img}
\end{figure*}

A point cloud of $50{,}000$ points is uniformly sampled from the aircraft mesh, and 98 primitives are then fitted, including 85 planes and 13 cylinders. To facilitate visualization, Fig.~\ref{fig:aircraft_prims} shows the fitted result in a segmented point-cloud form, where each color corresponds to one parametric primitive. 
All six engines are accurately detected as cylindrical primitives. 
In addition, cylindrical structures are also identified on the fuselage, nose, and wing leading edges, while the remaining regions are primarily fitted as planes. 
Ideally, wing edges should be fitted as several long and continuous planes. 
Although the wing edges are fragmented into several short planar patches due to limited sampling density, these structures contribute weakly to scattering and therefore have negligible influence on the subsequent modeling and imaging results.


\begin{figure}[t]
	\centering
	\includegraphics[width=1\linewidth]{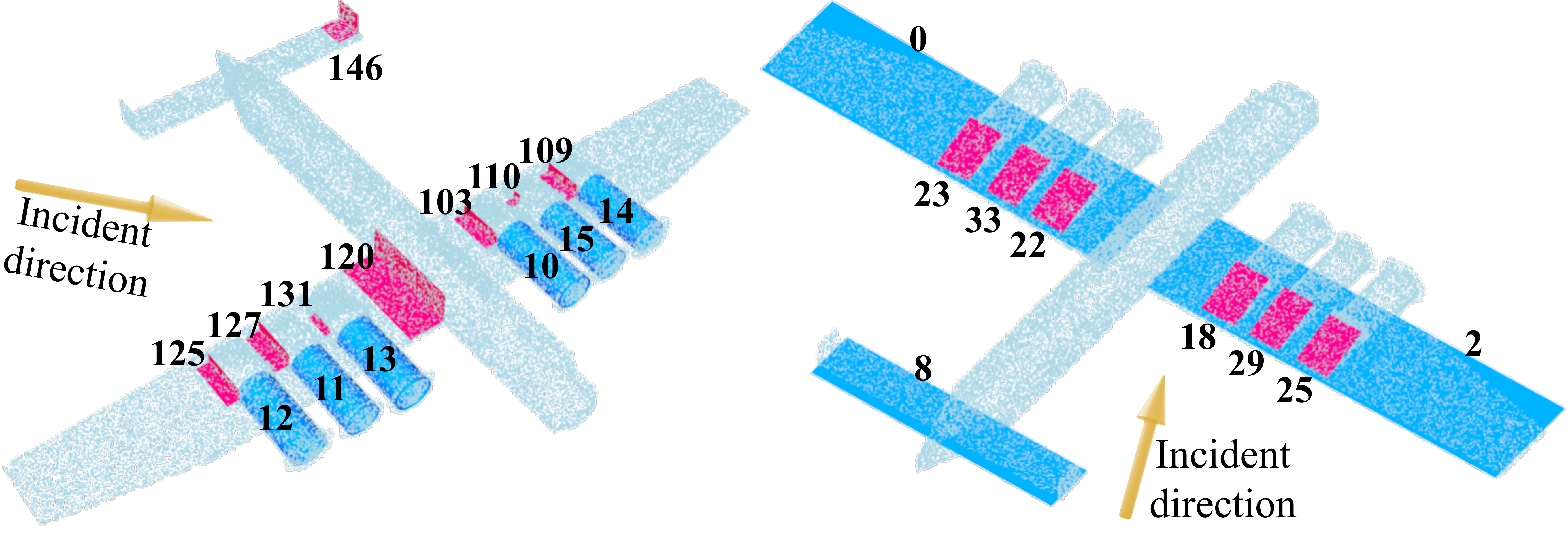}
	\caption{Dominant scatterers of the airplane target under fixed observation conditions. Left: the first column in Fig.~\ref{fig:aircraft_sar_img}, with 6 cylinders in blue and 8 dihedrals in red. Right: the second column in Fig.~\ref{fig:aircraft_sar_img}, with 9 planar scatterers highlighted in blue and red.}
	\label{fig:aircraft_scatterers}
\end{figure}

\begin{figure}[t]
	\centering
	\includegraphics[width=\linewidth]{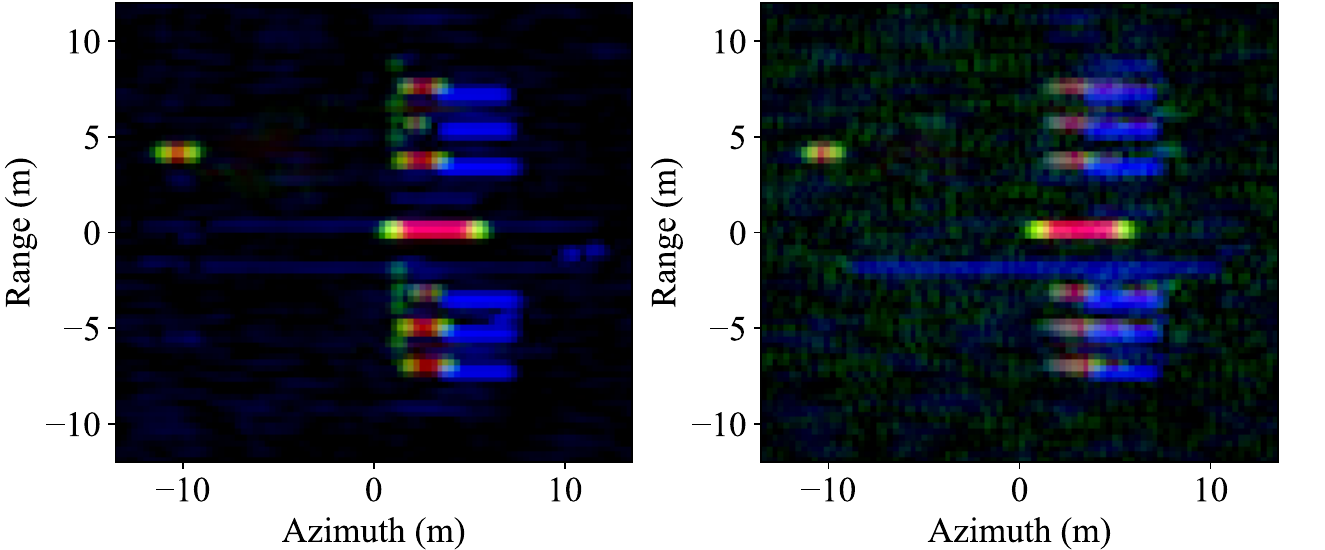}
	\caption{Comparison of full-polarimetric Pauli RGB pseudo-color images of the aircraft target generated by the proposed method (left) and FEKO (right).}
	\label{fig:aircraft_polsar}
\end{figure}

Based on the detected primitives, scatterers are further constructed. 
A total of 55 dihedrals are identified, resulting in 153 potential scatterers for the aircraft target. 

Fig.~\ref{fig:aircraft_sar_img} presents the simulated SAR images under three representative observation configurations, together with the FEKO results. 
In the first case, the range and azimuth resolutions are 0.15\,m and 0.1432\,m, respectively. 
Fourteen scatterers produce prominent responses. 
The strongest response corresponds to scatterer 120, which is a dihedral formed by the fuselage and the right wing. 
The second strongest response is scatterer 146, corresponding to a dihedral formed by the horizontal tail and the left vertical tail. 
The remaining 12 significant scatterers are contributed by the six engines, where each engine generates one cylindrical single-scattering (indices 10$\sim$15) and one dihedral formed by the rear vertical panel and the wing (indices 103, 109, 110, 125, 127, and 131).
The similarity between the two SAR images is $0.979$.
The effective regions of these dominant scatterers on the aircraft are visualized in Fig.~\ref{fig:aircraft_scatterers}, demonstrating that the proposed method can correctly detect potential multiple-bounce structures and extract physically meaningful effective regions through geometric clipping.

\begin{table}[t]
	\centering
	\caption{Scattering peak position matching results corresponding to the observation settings in Figs.~\ref{fig:slicy_sar}, \ref{fig:modi_slicy_sar_img}, \ref{fig:ship_sar_img}, and \ref{fig:aircraft_sar_img}.}
	\label{tab:peak_matching}
	\scriptsize
	\setlength{\tabcolsep}{3pt}
	\renewcommand{\arraystretch}{1.05}
	\begin{tabular}{@{}lcccccccccc@{}}
		\toprule
		& \multicolumn{3}{c}{SLICY} & Mod.SLICY & \multicolumn{3}{c}{Ship} & \multicolumn{3}{c}{Aircraft} \\
		\cmidrule(lr){2-4} \cmidrule(lr){5-5} \cmidrule(lr){6-8} \cmidrule(lr){9-11}
		& obs1 & obs2 & obs3 & obs1 & obs1 & obs2 & obs3 & obs1 & obs2 & obs3 \\
		\midrule
		\(R_{\mathrm{m}}\) & 1 & 1 & 0.875 & 0.75 & 1 & 0.9167 & 0.5 & 0.75 & 1 & 0.7692 \\
		\(P_{\mathrm{m}}\) & 0.875 & 1 & 1 & 1 & 1 & 0.8462 & 1 & 0.75 & 1 & 0.7692 \\
		\(E_{\mathrm{loc}}\) & 0 & 0 & 0.2857 & 0.8333 & 0 & 1.2042 & 1 & 1.5 & 0 & 1.4964 \\
		\bottomrule
	\end{tabular}
\end{table}

The second column of Fig.~\ref{fig:aircraft_sar_img} corresponds to a rear-looking configuration, for which the range and azimuth resolutions are both 0.2\,m.  
Under this geometry, nine scatterers dominate the SAR image. 
Two major scattering groups are produced by the rear panels of three engines on the left side (indices 23, 33, and 22) and on the right side (indices 18, 29, and 25), respectively, while the remaining weaker responses originate from the left wing (index 0), right wing (index 2), and horizontal tail (index 8). 
The resulting SAR image similarity with FEKO reaches 0.9965, and the spatial distributions of these dominant scatterers are visualized in Fig.~\ref{fig:aircraft_scatterers}.

The dominant scatterers under the observation configuration of the third column in Fig.~\ref{fig:aircraft_sar_img} are consistent with those in the first column. 
Although the locations and shapes of the scattering centers vary with the observation, the results remain highly consistent with the FEKO simulations, achieving a similarity of 0.9433. 

Fig.~\ref{fig:aircraft_polsar} compares the Pauli RGB images of the aircraft target under monostatic observation, with the incident frequency ranging from 0.75 to 1.25 GHz, an elevation angle of $35^\circ$, and an azimuth range of $-14.3^\circ$ to $14.3^\circ$. The proposed framework and FEKO show consistent bright scattering regions and color distributions, capturing the polarimetric responses from cylindrical and dihedral structures. In addition, the FEKO result contains a more evident weak-scattering background.

\subsection{Cross-Target Quantitative Evaluation}

\begin{figure}[t]
	\centering
	\includegraphics[width=\linewidth]{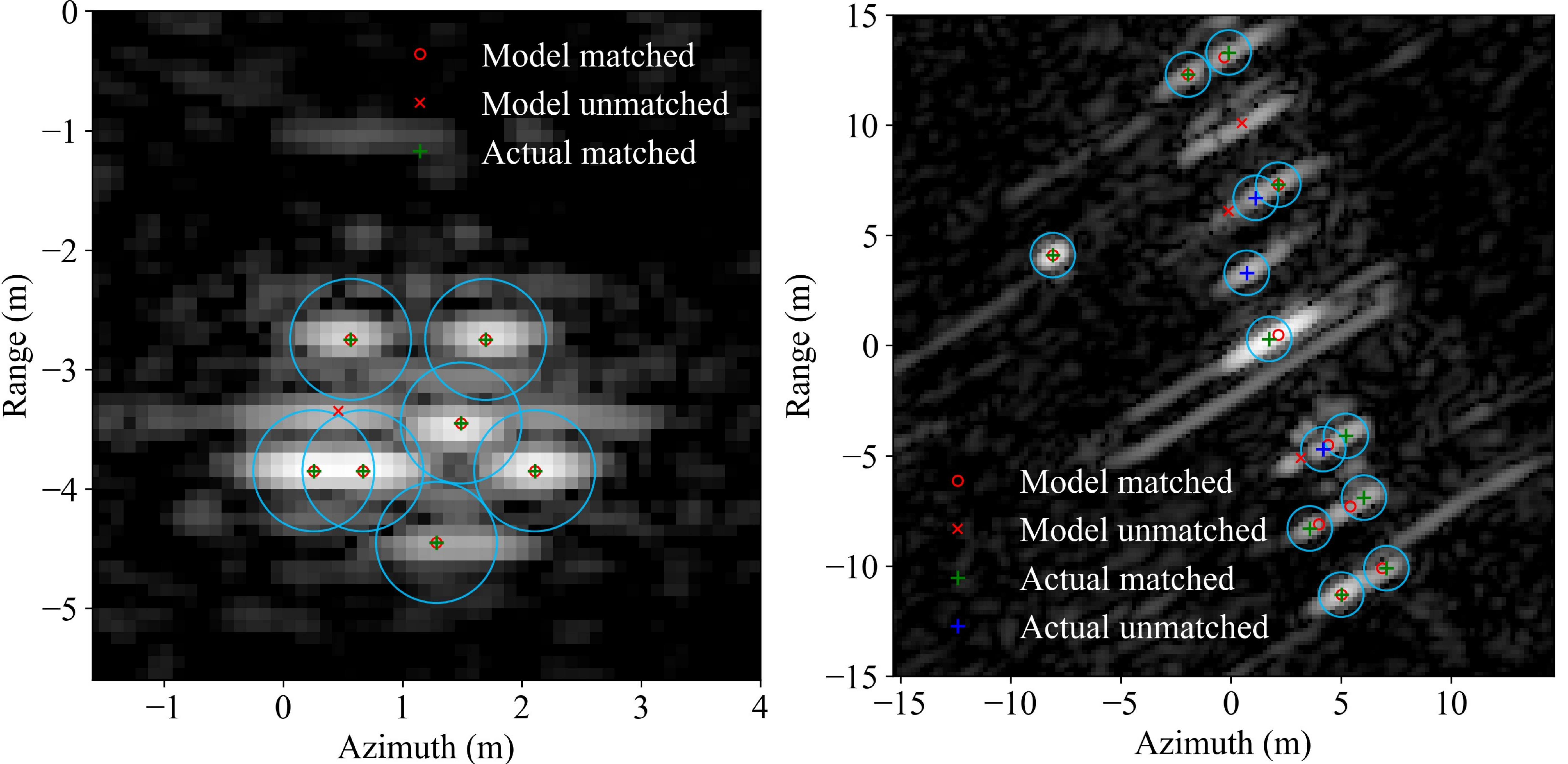}
	\caption{Visualization of scattering peak matching. The background is the reference SAR image, and the blue circles indicate the matching tolerance regions centered at reference peaks with a radius of 5.5 pixels. Left: SLICY-obs1; right: Aircraft-obs3.}
	\label{fig:peak_matching}
\end{figure}

\begin{table*}[t]
	\centering
	\caption{Runtime comparison between the proposed framework and FEKO RL-GO.}
	\label{tab:runtime}
	\scriptsize
	\setlength{\tabcolsep}{5.5pt}
	\renewcommand{\arraystretch}{0.8}
	\begin{tabular}{@{}lcccccccccc@{}}
		\toprule
		& \multicolumn{7}{c}{Proposed Framework}
		& \multicolumn{3}{c}{FEKO RL-GO} \\
		\cmidrule(lr){2-8} \cmidrule(lr){9-11}
		\multirowcell{2}{Target} &
		\multirowcell{2}{\makecell{Points}} &
		\multirowcell{2}{\makecell{Derivation\\Time}} &
		\multirowcell{2}{\makecell{Single\\Scatterers}} &
		\multirowcell{2}{\makecell{Multiple\\Scatterers}} &
		\multicolumn{3}{c}{Simulation Time} &
		\multicolumn{3}{c}{Simulation Time} \\
		\cmidrule(lr){6-8} \cmidrule(lr){9-11}
		& & & & &
		\(3\,\mathrm{GHz}\) Bi. & \(6\,\mathrm{GHz}\) Bi. & \(3\,\mathrm{GHz}\) Mono. &
		\(3\,\mathrm{GHz}\) Bi. & \(6\,\mathrm{GHz}\) Bi. & \(3\,\mathrm{GHz}\) Mono. \\
		\midrule
		SLICY & 50,000 & \(54.72\,\mathrm{s}+3.09\,\mathrm{s}\) & 19 & 14
		& \(0.054\,\mathrm{s}\) & \(0.053\,\mathrm{s}\) & \(0.052\,\mathrm{s}\)
		& \(4\,\mathrm{min}\,53\,\mathrm{s}\) & \(16\,\mathrm{min}\,42\,\mathrm{s}\) & \(6\,\mathrm{h}\,48\,\mathrm{min}\) \\
		
		Modified SLICY & 10,000 & \(3.20\,\mathrm{s}+0.09\,\mathrm{s}\) & 8 & 0
		& \(0.015\,\mathrm{s}\) & \(0.015\,\mathrm{s}\) & \(0.015\,\mathrm{s}\)
		& \(4\,\mathrm{min}\,2\,\mathrm{s}\) & \(13\,\mathrm{min}\,2\,\mathrm{s}\) & \(6\,\mathrm{h}\,12\,\mathrm{min}\) \\
		
		Ship & 50,000 & \(98.16\,\mathrm{s}+3.31\,\mathrm{s}\) & 30 & 16
		& \(0.070\,\mathrm{s}\) & \(0.069\,\mathrm{s}\) & \(0.069\,\mathrm{s}\)
		& \(1\,\mathrm{h}\,40\,\mathrm{min}\) & \(6\,\mathrm{h}\,13\,\mathrm{min}\) & \(>72\,\mathrm{h}\) \\
		
		Aircraft & 50,000 & \(1124.53\,\mathrm{s}+3.14\,\mathrm{s}\) & 98 & 55
		& \(0.219\,\mathrm{s}\) & \(0.221\,\mathrm{s}\) & \(0.234\,\mathrm{s}\)
		& \(2\,\mathrm{h}\,52\,\mathrm{min}\) & \(10\,\mathrm{h}\,22\,\mathrm{min}\) & \(>72\,\mathrm{h}\) \\
		\bottomrule
	\end{tabular}
	
	\vspace{0.6em}
	\begin{minipage}{0.99\textwidth}
		\scriptsize
		\textit{Note:} The proposed framework is tested on an Intel Core i5-13400F, and FEKO RL-GO is tested on a 16-core Intel Xeon Gold 6250. Derivation time includes primitive fitting and coupling detection. Simulation time is averaged over three random observation settings with \(1\,\mathrm{GHz}\) bandwidth and \(128\times128\) samples. Bi. = bistatic, Mono. = monostatic.
	\end{minipage}
\end{table*}

The previous subsections have evaluated geometric parameter accuracy and SAR image similarity. This subsection further summarizes peak matching and computational efficiency across targets.

Using the metrics in \eqref{eq:match_recall}--\eqref{eq:loc_error}, Table~\ref{tab:peak_matching} reports the peak matching results for all cases. Most cases achieve high recall and precision, with all \(E_{\mathrm{loc}}\) values within 1.5 pixels. Fig.~\ref{fig:peak_matching} shows two representative examples, where matched peaks generally correspond to semantically meaningful scattering structures, while unmatched peaks mainly appear in weak-scattering regions or near local sidelobes. In SLICY-obs1, all matched peaks are aligned at the pixel level, leading to \(E_{\mathrm{loc}}=0\).

Computational efficiency is another key advantage of the proposed framework. Table~\ref{tab:runtime} compares scatterer derivation time, including primitive fitting and coupling detection, and per-view simulation time with FEKO RL-GO. The proposed method concentrates most computation in the one-time scatterer construction stage; afterward, scattering responses under arbitrary views can be simulated in less than 1 s, with little dependence on frequency or observation mode. In contrast, RL-GO requires minutes to days for a single simulation, with cost increasing significantly with target complexity and frequency.

\subsection{Validation on Measured Data}
\label{sec:mstar_validation}

\begin{figure}[t]
	\centering
	\includegraphics[width=0.9\linewidth]{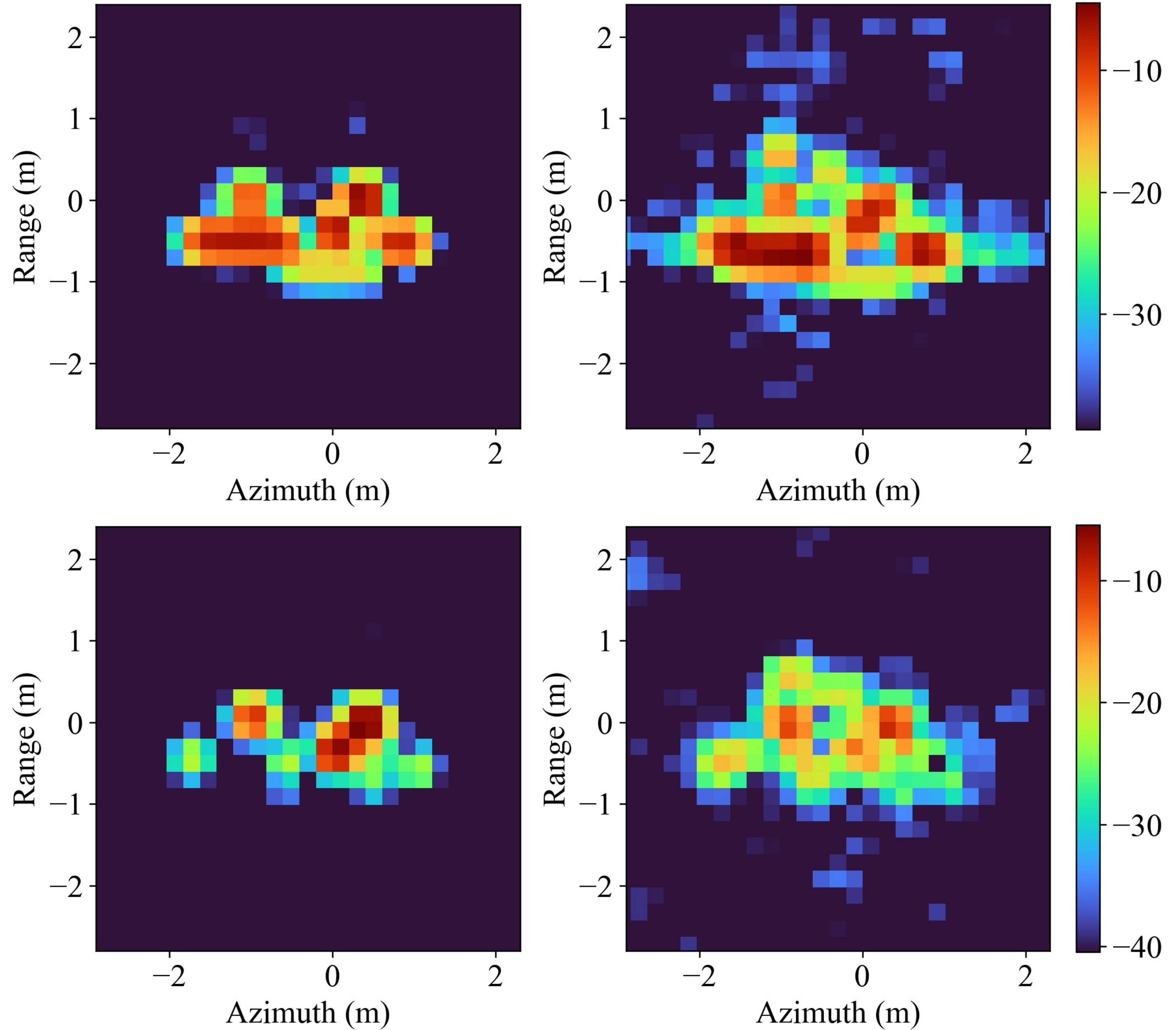}
	\caption{Comparison between simulated SAR images generated by the proposed method (left) and measured MSTAR images (right). The two rows correspond to center azimuth angles of \(0.952^\circ\) and \(3.952^\circ\), respectively.}
	\label{fig:slicy_sar_real}
\end{figure}

Finally, the SLICY model constructed above is further validated using measured data from the MSTAR dataset~\cite{Diemunsch1998MSTARATR}. 
The scatterers extracted from the aforementioned SLICY model are uniformly scaled to match the measured target size. Imaging parameters are taken from the MSTAR header, with a center frequency of \(9.599\,\mathrm{GHz}\) and a bandwidth of \(591\,\mathrm{MHz}\). Since the chip center may be shifted from the geometric-model origin, an overall translation is estimated on a common physical grid and compensated before comparison.

Fig.~\ref{fig:slicy_sar_real} compares simulated and measured SAR images at a depression angle of \(45^\circ\), with center azimuth angles of \(0.952^\circ\) and \(3.952^\circ\). When the azimuth angle is close to \(0^\circ\), the six scatterers analyzed above still dominate the response. An additional scattering center appears in the upper-left region of the measured image, possibly caused by cavity scattering from the open top of the short cylinder in the real SLICY target. As the azimuth angle slightly increases, the three originally distributed scattering centers disappear, leaving endpoint responses dominant. The image similarities for the two cases are 0.8274 and 0.8154, respectively.

\begin{figure*}[t]
	\centering
	\includegraphics[width=1\linewidth]{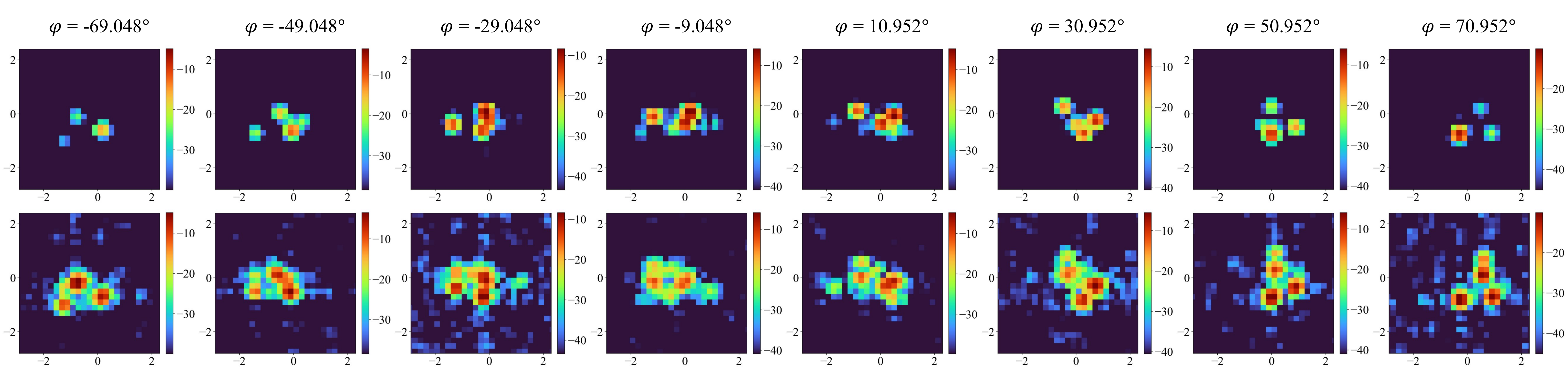}
	\caption{Simulated (top) and measured MSTAR (bottom) SLICY SAR images at different center azimuth angles and a depression angle of $45^\circ$.}
	\label{fig:slicy_sar_real2}
\end{figure*}

\begin{table}[t]
	\centering
	\caption{Quantitative comparison between simulated and measured SLICY SAR images over 303 center azimuth angles.}
	\label{tab:slicy_measured}
	\scriptsize
	\setlength{\tabcolsep}{10pt}
	\renewcommand{\arraystretch}{1.05}
	\begin{tabular}{lcc}
		\toprule
		Metric & Mean $\pm$ Std. & Median [P10, P90] \\
		\midrule
		$\mathrm{Cor}\uparrow$
		& $0.7689 \pm 0.0880$
		& $0.7798\ [0.6520,\,0.8767]$ \\
		
		$R_{\mathrm{m}}\uparrow$
		& $0.5712 \pm 0.2083$
		& $0.6000\ [0.2500,\,0.7500]$ \\
		
		$P_{\mathrm{m}}\uparrow$
		& $0.9168 \pm 0.1873$
		& $1.0000\ [0.6667,\,1.0000]$ \\
		
		$E_{\mathrm{loc}}\downarrow$ (pixel)
		& $0.6183 \pm 0.4506$
		& $0.5000\ [0.0000,\,1.2071]$ \\
		\bottomrule
	\end{tabular}
\end{table}

The agreement remains stable over a wide azimuth range. Fig.~\ref{fig:slicy_sar_real2} shows results from $-69.048^\circ$ to $70.952^\circ$ at $20^\circ$ intervals and a fixed depression angle of $45^\circ$. The scattering response varies substantially with azimuth. Despite broader responses and additional clutter in the measured images, the simulation reproduces the dominant scattering center locations and their evolution across viewing angles.

Table~\ref{tab:slicy_measured} summarizes the results over all 303 center azimuth angles spanning $0^\circ$--$360^\circ$. The peak-matching precision of 0.9168 is substantially higher than the recall of 0.5712, indicating that most predicted peaks have measured counterparts, whereas the measured images contain additional responses caused by clutter, nonideal structures, or unmodeled higher-order scattering. The mean and median localization errors are 0.6183 and 0.5 pixels, respectively, demonstrating accurate preservation of scattering center positions.

\section{Conclusion and Discussion}
\label{sec:conclusion}

This paper presents a fully geometry-driven parametric scattering modeling framework for complex targets.
Unlike existing approaches that perform \emph{view-dependent computation} by determining scattering structures through ray tracing, the proposed framework constructs stable, compact, and physically meaningful scattering representations directly from intrinsic target geometry, which can be reused across different viewing angles. 
Extensive experiments demonstrate that the proposed method achieves high consistency with EM simulations and measured MSTAR SAR images.
As for future works, efforts can be made in the following aspects:
\begin{enumerate}
	\item The canonical scattering center models adopted in this work are built on relatively simple primitives, which may limit their ability to represent more complex geometries and scattering mechanisms. 
	Extending the current framework to incorporate more fundamental scattering models is an important direction.
	
	\item For targets spanning multiple scales from global structures to fine details, primitive fitting can be sensitive to parameters, which may hinder the stable detection of small but important structures. 
	A hierarchical modeling strategy (e.g., a \emph{target--component--primitive} decomposition) could be explored to improve its robustness.

	\item The current method does not model global occlusion and may overestimate responses from occluded scatterers. Future work could introduce lightweight view-dependent visibility reasoning while preserving the reusable geometry-derived representation.
	
	\item This work focuses on forward scattering modeling. 
	An important future direction is to integrate the proposed framework with SAR inverse problems, such as scattering parameter inversion and target reconstruction, and to further exploit its clear semantics, compact parameterization, and physical interpretability.
\end{enumerate}

\small
\bibliographystyle{IEEEtran}
\bibliography{IEEEabrv,main}

\end{document}